\documentclass{jpp}
\usepackage[utf8]{inputenc}
\usepackage[T1]{fontenc}

\usepackage{amsmath}
\usepackage{amssymb}
\usepackage{graphicx}
\usepackage{bm}
\usepackage{mathrsfs}
\usepackage{color}
\usepackage{subfigure}
\usepackage{epstopdf, epsfig}
\usepackage{mathabx}
\usepackage{array} 
\usepackage{stackengine}
\usepackage{xcolor}

\usepackage{ulem}

\stackMath

\usepackage{natbib}
\newcommand{\vpar}{v_{\parallel}}
\newcommand{\beq}{\begin{equation}}
\newcommand{\eeq}{\end{equation}}

\newcommand{\bpar}{B_\parallel}
\newcommand{\apar}{ A_{\parallel}}

\newcommand{\kpe}{k_{\perp}^2}
\newcommand{\lapp}{\nabla_{\perp}^2}

\newcommand{\pa}{\partial}

\newcommand{\td}{\text{d}}

\newcommand{\nno}{\nonumber}

\newcommand{\bv}{\textbf{v}}

\newcommand{\br}{\bold{r}}
\newcommand{\bz}{\bold{z}}

\newcommand{\ben}{\begin{eqnarray}}
\newcommand{\een}{\end{eqnarray}}

\newcommand{\call}{\mathcal{L}}

\newcommand{\rs}{\rho_s}

\newcommand{\gamue}{G_{10e}}

\newcommand{\gamde}{G_{20e}}

\newcommand{\bee}{\beta_e}

\newcommand{\toe}{T_{0e}}

\newcommand{\lue}{\mathcal{L}_{U_e}}

\newcommand{\hfs}{g_s} 
\newcommand{\hv}{\hat{v}_\parallel}

\newcommand{\hcalf}{\hat{\mathcal{F}}_{{eq}_s}}

\newcommand{\hwu}{\hat{\mathcal{W}}}

\newcommand{\jo}{\mathcal{J}_{0s}}
\newcommand{\ju}{\mathcal{J}_{1s}}

\newcommand{\vts}{v_{{th}_s}}

\newcommand{\hvp}{\hat{v}_\perp}
\newcommand{\vp}{v_\perp}

\shorttitle{Plasmoid instability based on gyrofluid and gyrokinetic}
\shortauthor{C. Granier, R. Numata, D. Borgogno, E. Tassi, D. Grasso}

\title{Investigation of the collisionless plasmoid instability based on gyrofluid and gyrokinetic integrated approach}

\author{C. Granier \aff{1}
  \corresp{\email{camille.granier@oca.eu}},
  R. Numata\aff{2}
  D. Borgogno\aff{3}
 E. Tassi\aff{1}
 \and D. Grasso\aff{3}}

\affiliation{\aff{1}Universit\'e C\^ote d'Azur, CNRS, Observatoire de la C\^ote d'Azur, Laboratoire J. L. Lagrange, Boulevard de l'Observatoire, CS 34229, 06304 Nice Cedex 4, France
\aff{2}Graduate School of Information Science, University of Hyogo, Kobe 650-0047, Japan
\aff{3} Istituto dei Sistemi Complessi - CNR and Dipartimento di Energia, Politecnico di Torino, Torino 10129, Italy}

\begin{document}

\maketitle

In this work, the development of two-dimensional current sheets with respect to  tearing-modes, in collisionless plasmas with a strong guide field, is analysed. During their non-linear evolution, these thin current sheets can become unstable to the formation of plasmoids, which allows the magnetic reconnection process to reach high reconnection rates. We carry out a detailed study of the impact of a finite $\beta_e$, which also implies finite electron Larmor radius effects, on the collisionless plasmoid instability. This study is conducted through a comparison of gyrofluid and gyrokinetic simulations. The comparison shows in general a good capability of the gyrofluid models in predicting the plasmoid instability observed with gyrokinetic simulations. We show that the effects of $\beta_e$ promotes the plasmoid growth. The impact of the closure applied during the derivation of the gyrofluid model is also studied through the comparison of the energy variation.

\section{Introduction}

Magnetic reconnection is a change of topology of the magnetic field lines taking place in  regions of intense localized current, referred to as current sheets. This fundamental process ultimately converts magnetic energy into bulk flow and particle heating, and is responsible for the explosive release of magnetic energy in astrophysical and laboratory plasmas. The instabilities of very elongated reconnecting current sheets leading to the formation of secondary magnetic islands, called plasmoids, have generated a lot of interest, as they are believed to achieve fast reconnection. Plasmoids have been greatly studied through the most standard reconnection model based on the Sweet-Parker (SP) theory in the resistive magnetohydrodynamics (RMHD) framework (\cite{Bis96, Lou05}). 
In \cite{Bis96}, it has been shown that collisional current sheets become unstable when $S= \mu_0 L_{cs} v_A /\eta > 10^4$, where $L_{cs}$ is the length of the current sheet, $\eta$ is the resistivity and $v_A$ is the Alfvén speed. 
Much work has followed and allowed to identify the plasmoid regime as a function of the Lundquist number $ S $ and of the characteristic scale of a dynamic of the ions (ion-sound Larmor radius $\rs$ or ion skin depth $d_i$ scales) at which a transition to a non-collisional regime, dominated by kinetic effects, occurs (\cite{Ji11, Dau12, Lou15, Uzd10, Uzd15, Bha18}). 
This extension of the resistive reconnection regime with the inclusion of the ion dynamics enlarged the study to a broader parameter space, but also suggested that plasmoids are fundamental features of reconnecting  current sheets, regardless the value of the Lundquist number (\cite{Ji11, Dau12}).

The plasma in the magnetosphere and solar wind, which regularly undergoes reconnection, is so dilute that collisions between particles are extremely infrequent. In such plasmas, electron inertia becomes particularly relevant to drive reconnection in thin current sheets.
Indeed, recent observations revealed many reconnection onsets driven by electrons, in the presence of a strong guide field, close to the dayside magnetopause and magnetosheath \cite{Bur16,Pha18} with current sheets having a thickness of the order of the electron inertial length.  Regarding experiments, a study by \cite{Ols16} also gave direct experimental proof of plasmoid formation at the electron scale in a weakly collisional regime. 
 In these collisionless, magnetized environments, effects of the finite electron Larmor radius (FLR) on the reconnection process can also become non-negligible, in particular when $\beta_e$, being defined as the ratio between equilibrium thermal electron pressure and  guide field magnetic pressure, is not much smaller than unity. This motivates the study of the formation of plasmoids in non-collisional current sheets, and in particular, the impact of the effects relevant at the electron scales such as the electron skin depth and the electron Larmor radius. 

In \cite{Gra2022pl} instability thresholds for the purely collisionless plasmoid onset were clearly identified in the regime $\beta_e \rightarrow 0$.  In this article we relax the assumption of small $\beta_e$ and carry out a detailed study of the impact of a finite $\beta_e$,  on the collisionless plasmoid instability, in the case of a strong guide field. We consider inertial reconnection, and finite electron FLR effects arise from the combination of electron inertia and finite $\beta_e$ parameters. This study is conducted through a comparison of gyrofluid and gyrokinetic simulations. Both approaches are assuming that the plasma is immersed in a strong guide field directed along the $z$ direction. As a by-product of our analysis, we also obtain a way to validate, by means of gyrokinetic simulations, part of  the results on collisionless plasmoid instability obtained by \cite{Gra2022pl} with a gyrofluid approach in the $\beta \rightarrow 0$ limit (later referred to as {\it fluid} limit) 

The gyrofluid model is the 2-field system presented in (\cite{Gra22flr}) and assumes cold and immobile ions along the guide field direction.  Gyrofluid models, although greatly simplified with respect to the original gyrokinetic system, are useful tools for studying collisionless reconnection, in which the microscopic scales, such as the electron skin depth and the electron Larmor radius, can be more important than resistivity. In addition, the gyrofluid framework is less costly in terms of computational resources, and physically more intuitive when compared to the kinetic or gyrokinetic framework. So far, gyrofluid modelling allowed to gain a good understanding of the role of collisionless effects (e.g. \cite{Del11, Com13, Tas18, Gra21, Gra22flr}). 

The gyrokinetic model, adopted for the comparison, is a $\delta f$ model which solves the electromagnetic gyrokinetic Vlasov-Maxwell system. The gyrokinetic equations are solved by means of the {\tt AstroGK} code, presented and used in \cite{Num10, Num15}.  One of the main advantages of using the {\tt AstroGK} code for  a comparison with the gyrofluid results, is that, in a specific limit, the gyrokinetic system solved by {\tt AstroGK} reduces to the one   that was taken to derive the 2-field gyrofluid model used in this study \citep{How06}. This allows to study the impact of the closure applied on the moments, performed during the derivation of the gyrofluid model, on the distribution and conversion of energy during reconnection and identify the possible limitations of the gyrofluid approach.  The specific limit in which the {\tt AstroGK} code has to be used, in order to reproduce the parent gyrokinetic model of the gyrofluid model, is that corresponding to a straight guide field, with no density and temperature gradients and without collisions. To be consistent with the gyrofluid approach, the ions are assumed to be cold.

The article is organized as follow. In Sec. 2 we present the gyrofluid and gyrokinetic systems, as well as the numerical set up. In Sec. 3 we present the results concerning the plasmoid instability obtained from a comparison of the two approaches. In Sec. 4 we compare the energy variations in the two frameworks and discuss the impact of the closure hypothesis on the energy conversion. Section 5 is devoted to conclusions.

\section{Adopted models}
\subsection{Gyrofluid}
The gyrofluid model used for our analysis is the one considered by \cite{Gra22flr}, which consists of the following evolution equations
\begin{align}
&\frac{\pa N_e}{\pa t}+[\gamue \phi - \rs^2 2 \gamde \bpar , N_e]- [\gamue \apar , U_e]=0,  \label{conteiso}\\
&\frac{\pa A_e}{\pa t}+[\gamue \phi - \rs^2 2 \gamde \bpar , A_e]+\rs^2[\gamue \apar ,N_e]=0,  \label{momeiso}
\end{align}
complemented by the relations 
 \begin{align}
&\left( \frac{ \gamue^2 -1}{\rs^2}+\lapp\right) \phi-\left(  \gamue 2 \gamde -1\right)\bpar =\gamue N_e,  \label{qncondiso}\\
&\lapp\apar=\gamue U_e,  \label{ampparcondiso}\\
&\left( \gamue 2 \gamde-1\right)\frac{\phi}{\rs^2} -\left(\frac{2}{\bee}+ 4 \gamde^2\right)\bpar=2 \gamde N_e.  \label{ampperpcondiso}
\end{align}
Equation (\ref{conteiso}) corresponds to the electron gyrocenter continuity equation, whereas Eq. (\ref{momeiso}) refers to the electron momentum conservation law, along the guide field direction.
The static relations (\ref{qncondiso}), (\ref{ampparcondiso}) and (\ref{ampperpcondiso}) descend from quasi-neutrality and from the projections of Amp\`ere's law along directions parallel and perpendicular to the guide field, respectively. The guide field is directed along the $z$ axis of a Cartesian coordinate system $x,y,z$, and, in the present 2D version of the model, the dependent variables are functions only of $x$ and $y$, as well as of the time variable $t$. We indicated with $N_e$ and $U_e$ the fluctuations of the electron gyrocenter density and parallel velocity, respectively, whereas $\phi$ and $\bpar$ indicate the fluctuations of the electrostatic potential and of the magnetic field along the guide field. The variable $A_e$ is defined by $A_e=\gamue \apar - d_e^2 U_e$, where $\apar$ is the $z$-component of the magnetic vector potential, $d_e=\sqrt{m_e c^2/{4 \pi e^2 n_0}}/L$ is the normalized electron skin depth and $\gamue$ is an electron gyroaverage operator, defined later in Eq. (\ref{gyroave}).    The operator $[ \, , \, ]$ is the canonical Poisson bracket and is defined by $[f,g]=\partial_x f \partial_y g - \partial_y f \partial_x g$, for two functions $f$ and $g$. The perpendicular Laplacian operator $\lapp$ is defined by $\lapp f=\partial_{xx}f + \partial_{yy} f$. The variables are normalized as
\begin{align}
& t=\frac{v_A}{L}\hat{t}, \qquad x=\frac{\hat{x}}{L}, \qquad y=\frac{\hat{y}}{L}, \label{norm1} \\
& d_{i} N_{e,i}= \frac{\hat{N}_{e,i}}{n_0}, \qquad d_i U_{e,i}=\frac{\hat{U}_{e,i}}{v_A}, \label{norm2}\\
& \apar=\frac{\hat{A}_\parallel}{L B_0}, \qquad d_i \bpar= \frac{\hat{B}_\parallel}{B_0}, \qquad \phi=\frac{c}{v_A} \frac{\hat{\phi}}{L B_0}, \label{norm3}
\end{align}

where the hat indicates dimensional variables, $c$ is the speed of light, $L$ is a characteristic scale length, $n_0$ is the equilibrium uniform density,  $B_0$ is the amplitude of the guide field and $v_A=B_0/\sqrt{4 \pi m_i n_0}$ is the Alfv\'en speed, with $m_i$ indicating the ion mass. The normalized ion skin depth is defined by $d_i=\sqrt{m_i c^2/{4 \pi e^2 n_0}}/L$, where $e$ indicates the proton charge. In Eq. (\ref{norm2}) we also introduced the quantities $N_i$ and $U_i$, corresponding to the ion gyrocenter density and parallel velocity fluctuations, respectively. Such moments do not  evolve in the model (\ref{conteiso})-(\ref{ampperpcondiso}), and the assumptions on such quantities will be discussed later in this section, as well as in Sec. \ref{ssec:compar}. We find it also useful to write explicitly  the expression for the magnetic field normalized with respect to the guide field amplitude. In the present 2D setting, by virtue of the normalization (\ref{norm1})-(\ref{norm3}), such expression is given by 
\begin{equation} \label{magfield}
   \mathbf{B}(x,y,t) =  \mathbf{z}+ d_i \bpar(x,y,t)\mathbf{z} + \nabla \apar (x,y,t)\times \mathbf{z},
\end{equation}
where $\mathbf{z}$ is the unit vector along $z$.
 Independent parameters in the model are $\bee=8 \pi n_0 \toe/B_0^2$, $\rs=\sqrt{{\toe m_i c^2}/{(e^2 B_{0}^2)}}/L$\footnote{According to a customary notation, in the symbols $\rs$, the subscript $s$ is to indicate a sonic quantity and not the particle species.} and $d_e=$, where $\toe$ is the uniform  equilibrium electron temperature and $m_e$ is the electron mass. These three parameters  correspond to the ratio between equilibrium electron pressure  and magnetic guide field pressure, to the normalized sonic Larmor radius and to the electron skin depth, respectively. 
 
 The model is formulated on a domain $\{ (x,y) \, : \, -L_x \leq x \leq L_x , -L_y \leq y \leq L_y \}$, with $L_x$ and $L_y$ positive constants. Periodic boundary conditions are assumed. This allows to express gyroaverage operators in terms of the corresponding Fourier multipliers. In particular, we associate the electron gyroaverage operators $\gamue$ and $\gamde$ with corresponding Fourier multipliers in the following way \citep{Bri92} 
 \begin{equation}   \label{gyroave}
  \gamue=2\gamde \rightarrow \mathrm{e}^{-\kpe \frac{\bee}{4}d_e^2},
\end{equation}
where $\kpe=k_x^2 + k_y^2$ is the squared perpendicular wave number and $k_x=m \pi/L_x$, $k_y=n \pi/L_y$ are the $x$ and $y$ components of the wave vector, with $m$ and $n$ positive integers.
As is customary with gyrofluid models, Eqs. (\ref{conteiso}) and (\ref{momeiso}) are expressed in terms of gyrocenter variables. However, for the sake of the subsequent analysis, it can be useful also to express their relation with particle variables. Such relation, in particular, is affected by the quasi-static assumption, used in the derivation of the model \citep{Tas20} to obtain a closure on the infinite hierarchy of moment equations obtained from a parent gyrokinetic system. As a consequence of such quasi-static closure (which will be briefly recalled in Sec. \ref{ssec:compar}) the normalized density fluctuations and parallel velocity fluctuations of the electrons, indicated with $n_e$ and $u_e$, respectively, are related to those of the corresponding gyrocenters by
\begin{align}
&  N_{e}= \gamue^{-1} \left( n_e + \left( \gamue^2 -1 \right) \frac{\phi}{\rs^2} - \gamue^2 \bpar \right),  \label{ne} \\
&  U_{e}= \gamue^{-1} u_e. \label{ue}
\end{align}
Also, in our gyrofluid model we neglect the contributions due to the density and parallel velocity fluctuations of the ion gyrocenters, by imposing that $N_i=0$, $U_i=0$. Furthermore, ions are assumed to be cold, i.e. $\tau \rightarrow 0$, where $\tau=T_{0i}/\toe$ is the ratio between ion and electron equilibrium temperature.

In terms of the ion particle density and parallel velocity fluctuations, denoted as  $n_i$ and $u_i$, respectively, such assumptions lead to the relations $n_i= \lapp  \phi + \bpar=  n_e$ and $u_i=0$.  
From the quasi-neutrality relation (\ref{qncondiso}), Amp\`ere's law (\ref{ampparcondiso})-(\ref{ampperpcondiso}), combined with Eqs. (\ref{ne})-(\ref{ue}), we can obtain the relations
\begin{align}
&  n_e = \frac{2}{2 + \bee} \lapp \phi= - \frac{2}{\bee} \bpar,\label{ne2} \\
&  u_e = \lapp \apar, \label{ue2}
\end{align}
that permit to express the electron particle (as opposed to gyrocenter) density and parallel velocity fluctuations, in terms of electromagnetic perturbations such as $\phi$, $\bpar$ and $\apar$.

It is also particularly relevant to consider the limit $\bee \rightarrow 0$ with $d_e$ and $\rho_s$ remaining finite (which implies $m_e /m_i \rightarrow 0$). This corresponds to suppressing the effects of parallel magnetic perturbations and electron FLR effects. One of the purposes of our investigation is indeed to consider possible modifications, due to kinetic effects, of the plasmoid instability scenario described by Ref. \cite{Gra2022pl} and which was conceived namely in the regime with $\bee \rightarrow 0$ and finite $d_e$ and  $\rho_s$. In this limit, the gyroaverage operators can be approximated in the Fourier space in the following way:
 \begin{equation} \label{gyroop}
\gamue f(x,y) = 2 \gamde f(x,y) =f(x,y) + O(\bee).
\end{equation}
Using this development in Eqs. (\ref{conteiso}) - (\ref{ampperpcondiso}) and neglecting the first order corrections, we obtain the evolution equations (\cite{Sch94})
\begin{equation} \label{fluid1}
  \frac{\partial n_e}{\partial t} + [\phi, n_e] - [ \apar, u_e] = 0,
\end{equation} 
\begin{equation} \label{fluid2}
\begin{split} 
\frac{\partial}{\partial t} \left( \apar - d_e^2 u_e\right) + \left[\phi , \apar - d_e^2  u_e\right] - \rho_s^2[n_e, \apar] =0,
\end{split}
\end{equation}\\
where the static relations (\ref{qncondiso}) - (\ref{ampperpcondiso}) are replaced by 
\begin{align}
& \lapp \phi = N_e =n_e,  \label{qncondisof}\\
& \lapp\apar= U_e= u_e,  \label{ampparcondisof}\\
& \bpar=0. 
\end{align}
In this limit, particle density and parallel velocity fluctuations coincide with the corresponding gyrocenter counterparts. The system (\ref{fluid1})-(\ref{fluid2}), complemented by the static relations (\ref{qncondisof})-(\ref{ampparcondisof}) corresponds to the model by \cite{Caf98}, adopted  for describing collisionless reconnection in the presence of  electron inertia and finite sonic Larmor radius effects. Because of the absence of FLR effects, we will refer to the model (\ref{fluid1})-(\ref{ampparcondisof}) as to the {\it fluid} limit of the general gyrofluid model (\ref{conteiso})-(\ref{ampperpcondiso}).

\subsection{Gyrokinetic}   \label{ssec:gyrokin}
 

In this section, we present the electromagnetic $\delta f$ gyrokinetic model used in this work~\citep{How06,Num10}, from which the gyrofluid model can be derived with appropriate approximations and closure hypotheses \cite{Tas20}. 
The gyrokinetic model is formulated in terms of the perturbation of the gyrocenter distribution function 
$g_s = g_s(\textbf{X}_s, \vpar, \vp, t)$ where $\vpar$, $\vp$ are the parallel and perpendicular velocity coordinates. The guiding center coordinates is given by
\begin{equation}
\textbf{X}_s =\mathbf{x} + \frac{\vts}{\omega_{cs}} \bv\times \bz,
\end{equation}
where $\mathbf{x}$ is the particle position, $\bv$ is the particle velocity, $\vts=\sqrt{T_{0s}/m_s}$ is the thermal speed and $\omega_{cs}=e B_0 /(m_s c)$ is the cyclotron frequency.
The index $s$ labels the particle species, with $s=e$ for electrons and $s=i$ for ions. 
For simplicity, we assume a uniform background plasma, and two-dimensionality ($\partial /\partial z = 0)$. By adopting the same normalization scheme with the gyrofluid model, the gyrokinetic system can be written in the following way,

\begin{align}
& \frac{\pa }{\pa t}\left(g_s + \frac{1}{\rho_s} \frac{ Z_s}{\sqrt{\sigma_s\tau_s}} \vpar \jo \apar\right) +  \left[ \jo \phi - \sqrt{\frac{\tau_s \beta_e}{2\sigma_s}} \vpar \jo \apar + \frac{\tau_s}{Z_s} \rho_s^2 \vp^2 \ju \bpar , \hfs + \frac{1}{\rho_s} \frac{Z_s}{\sqrt{\sigma_s \tau_s}} \vpar \jo \apar \right] = 0, \label{gyr} \\
& \sum_{s} Z_{s} \bar{n}_s = \phi \sum_{s} \frac{Z_s}{\rho_s^2 \tau_s}
    \left( \frac{1}{n_0} \int \td \hwu \hcalf (1 - \jo) \right) + \bpar \sum_{s} Z_s \left(\frac{1}{n_0} \int \td \hwu \hcalf  v_\perp^2 \jo \ju \right), \label{qn} \\  
& \sum_{s} Z_{s} \bar{u}_s = - \nabla_{\perp}^{2} \apar,
\label{amp} \\
& \sum_{s} \bar{p}_s = - \phi \sum_{s} \frac{1}{\rho_s^2}\left(\frac{1}{n_0} \int \td \hwu \hcalf v_\perp^2 \jo \ju \right) - \bpar \left( \frac{2}{\beta_e} + \sum_{s} \int \td \hwu \hcalf ( v_\perp^2 \jo \ju)^2 \right) 
    \label{amppe}
\end{align}

Eq. (\ref{gyr}) is the gyrokinetic equation, whereas Eqs. (\ref{qn}), (\ref{amp}) and (\ref{amppe}) correspond to the quasi-neutrality relation and to the parallel and perpendicular projection of Amp\`ere's law.
We have introduced the following additional normalizations
\begin{align}
    g_s = \frac{\hat{g}_s}{\hcalf}, \qquad v_{\parallel,\perp} = \frac{\hat{v}_{\parallel,\perp}}{\vts},
    \qquad d_i \bar{p}_s = \frac{\hat{\bar{p}}_s}{n_0 T_{0e}}
\end{align}
where the Maxwellian equilibrium distribution function in the dimensional form is
\begin{equation}
\label{max}
\hcalf(\hv,\hvp)= n_0 \left(\frac{m_{s}}{{2 \pi T_{0s}}}\right)^{3/2} \mathrm{e}^{-\frac{m_{s} \hv^2}{2 T_{0s}}-\frac{m_s \hvp^2}{ 2 T_{0s} }}.
\end{equation}
The species dependent parameters are the mass ratio $\sigma_s = m_s/m_i$, temperature ratio $\tau_s = T_{0s}/T_{0e}$, and charge number ratio $Z_s = q_s/q_i$. Obviously, $\sigma_i=1$, $\tau_e = 1$, $Z_i=1$. We fix $Z_e=-1$, i.e. $q_i=e$, throughout this work. We may occasionally denote the non-trivial ones as $\sigma_e = \sigma$, $\tau_i = \tau$.

The velocity moments of the distribution function appear in Eqns.~\eqref{qn}-\eqref{amppe} are defined by
\begin{align}
    d_i \bar{n}_{s} & = \frac{1}{n_{0}} \int \td \hwu \hcalf \jo g_{s} , \\
    d_i \bar{u}_{s} & = \sqrt{\frac{\tau_s \beta_e}{2\sigma_s}} \frac{1}{n_{0}}
    \int \td \hwu \hcalf \vpar \jo g_{s},  \\
    d_i \bar{p}_{s} & = \tau_{s} \frac{1}{n_{0}} \int \td \hwu \hcalf \vp^2 \ju g_{s},
\end{align}
where the volume element $d \hwu$ in velocity space is defined as $d \hwu = \pi \vts^3 d \vpar d \vp^2$. Note that these quantities are moments of $g_s$, thus are different from the actually particle density, flow, and pressure, i.e. the moments of the total distribution function $\delta {\mathcal F}_s$ defined later on.

Finally, the gyroaverage operators $\jo$ and $\ju$ can be expressed, analogously to Eq. (\ref{gyroave}), in terms of Fourier multipliers in the following way:
\begin{align}
&\jo \rightarrow J_0 \left(\alpha_s \right), \qquad \ju \rightarrow \frac{J_{1}(\alpha_s)}{\alpha_s},
\end{align}
where $J_0$ and $J_1$ are the zeroth and first order Bessel functions, respectively, and the argument is defined by $\alpha_s = k_{\perp} \vp \vts/(L \omega_{cs}) = k_{\perp} \vp (\rho_s \sqrt{\sigma_s \tau_s}/Z_s)$.

\subsection{Connection between the gyrofluid and the gyrokinetic models} \label{ssec:connect}

We assume the distribution function can be written as
\beq
\hat{g}_s(\textbf{X}_s,\vpar,\vp,t) =
\hcalf \sum_{n,m=0}^{\infty}
    H_{m}(\vpar) L_{n}\left(\frac{\vp^2}{2}\right)
    f_{mn_{s}}(\textbf{X}_s,t)
\eeq
where $H_{m}$ and $L_{n}$ indicate the Hermite and Laguerre polynomials, respectively,  of order $m$ and $n$, with $m$ and $n$ non-negative integers. From the orthogonality properties of the Hermite and Laguerre polynomials, the following relation holds:
\beq   \label{moments}
f_{{mn}_s}=\frac{1}{n_0 \sqrt{m!}}\int \td \hwu \, \hcalf g_s H_m\left(\vpar\right) L_n\left(\frac{\vp^2}{2} \right).
\eeq
The functions $f_{mn_{s}}$ are coefficients of the expansion and  are proportional to fluctuations of the gyrofluid moments. Indeed, for instance, $f_{{20}_e}$ is proportional to gyrocenter electron parallel temperature fluctuations.

 In the present 2D case with an isotropic equilibrium temperature, the system is closed by a closure called "quasi-static" which was derived in \cite{Tas20} and which implies that, with the exception of $N_{e,i}$ and $U_{e,i}$,  all the other gyrofluid moments  are constrained by the relations

\begin{equation}
\label{qsmoments}
f_{{mn}_s} = - \delta_{m0} \left(
G_{1ns} \frac{1}{d_i}\frac{2Z_s}{\tau_s\beta_e} \phi + 2 G_{2ns} d_i \bpar
\right)
\end{equation}

where $\delta_{m0}$ is a Kronecker delta and $m$ and $n$ are non-negative integers, with $(m,n) \neq (0,0)$ and $(m,n) \neq (1,0)$, namely to exclude $N_{e,i}$ and $U_{e,i}$.
In Eq. (\ref{qsmoments}), we also introduced the gyroaverage operators which, as Fourier multipliers, are given by
 \begin{align}
 & G_{1n} (b_s) \rightarrow  \frac{\mathrm{e}^{-b_s /2}}{n!}\left(\frac{b_s}{2}\right)^n, \qquad n\geq 0,   \label{G1s}\\
 & G_{20} (b_s) \rightarrow \frac{\mathrm{e}^{-b_s /2}}{2}, \quad G_{2n} (b_s) \rightarrow -\frac{\mathrm{e}^{-b_s/2}}{2} \left( \left(\frac{b_s}{2}\right)^{n-1} \frac{1}{(n-1)!}-\left(\frac{b_s}{2}\right)^{n} \frac{1}{n !}\right), \qquad n \geq 1,  \label{G2s}
 \end{align}
 %
with $b_s=k_\perp^2 \vts^2/(L\omega_{cs})^2 = k_\perp^2 d_i^2 (\sigma_s \tau_s \beta_e/(2 Z_s^2))=k_{\perp}^2\rho_s^2 (\sigma_s\tau_s/Z_s^2)$.
Expressed in terms of particle variables, this closure implies that, with the exception of $n_{e,i}$ and $u_{e,i}$, the fluctuations of the particle moments are zero. 

The (2D) quasi-static closure is valid when 
 \beq  \label{qstatic2d}
 \frac{\hat{\omega}}{\hat{k}_y} \ll \vts 
 \eeq
 for $s=e,i$ are satisfied, where $\hat{k}_y$ is the $y$ component of the wave vector and $\hat{\omega}$ is the frequency obtained from the dispersion relation of the gyrokinetic equation (\ref{gyr}) linearized about an equilibrium  ${\phi}^{(0)}={B}_{\parallel}^{(0)}=0,A_{\parallel}^{(0)}=a x$, with a constant $a$. 
 Such linearization assumed that one can identify, also in real space, the Fourier multiplier $-k_\perp^2$ with the operator $\nabla_\perp^2$ and then make use of the identity $J_0 (\alpha_s)=\sum_{r=0}^{\infty}(1/(r!)^2)(\vp \vts/(L \omega_{cs}))^{2r}{\nabla_\perp^{2r}}$, so that $\jo x \approx x$.The condition (\ref{qstatic2d}) is better fulfilled for waves with small phase velocity along $y$, which justifies the term quasi-static.
 With regard to the moments not fixed by the quasi-static closure, we have that the dynamics of $N_e$ and $U_e$ is governed by the evolution equations (\ref{conteiso})-(\ref{momeiso}), that can then be obtained from the zeroth and first order moment, with respect to the parallel velocity coordinate $\vpar$, of the electron gyrokinetic equations (\ref{gyr}), with $s=e$.
 
 Concerning $N_i$ and $U_i$,  as above stated, we assume the conditions
 \beq  \label{ioncond}
 N_i=0, \qquad U_i=0
 \eeq
 to hold. This assumption effectively decouples the ion gyrofluid dynamics from the electron gyrofluid dynamics, leaving Eqs. (\ref{conteiso})-(\ref{ampperpcondiso}) a closed system.
 
 The assumption (\ref{ioncond}), as can be derived from the four-field model in \cite{Gra22flr} is valid for $\beta_e \ll 1$.\footnote{With this regard it could be useful to mention here a misprint in Eq. (2.2) in \cite{Gra22flr}, where $U_i$ should have been multiplied by the factor $2 \rho_{s_\perp}^2/ \beta_{e_\perp}$.} Therefore we expect, in the gyrokinetic simulations, a departure from the condition (\ref{ioncond}) as $\beta_e$ increases.
 
 We finally mention that in both gyrofluid and gyrokinetic simulations, we consider the cold-ion case, i.e.
 \beq
 \tau \ll 1,
 \eeq
 where we recall that $\tau =\tau_i = T_{0i}/T_{0e}$.

\subsection{Numerical set-up}

We assume an equilibrium in which the electromagnetic quantities are given by 
\beq   \label{equilgf}
\phi^{(0)} = 0, \quad \quad \apar^{(0)} =   A_{\parallel 0}^{eq} /\cosh^2 \left( x \right), \quad \quad \bpar^{(0)}=0,
\eeq
with $A_{\parallel 0}^{eq}=1.299$, in order to have $\mathrm{max}_x(B_{y}^{eq}(x))=1$. Due to periodicity assumption in the simulations\footnote{In the {\tt AstroGK} code, a shape function $S_h(x)$ is multiplied to $A_{\parallel}^{(0)}$ to enforce periodicity~\citep{Num10}. This minor difference in the simulation set-up between two models practically introduces no difference in the following results.}, the dimensionless equilibrium $\apar^{(0)}$ is replaced by 
\beq
 \apar^{(0)}=\sum_{n=-30}^{n=30}  A_{\parallel 0}^{eq} a_n \mathrm{e}^{in 2\pi x}, \qquad 
\eeq
 where $a_n$ are the Fourier coefficients of the function $1/\cosh^2 \left( x \right)$.  
 
 Note that, in order to satisfy Eqs. (\ref{qncondiso})-(\ref{ampperpcondiso}) at equilibrium, the corresponding equilibrium density and parallel electron velocity fluctuations of the electron gyrocenters are fixed as
 \beq  \label{eqneue}
 N_e^{(0)}=0, \qquad \lapp \apar^{(0)}=\gamue U_e^{(0)}.
 \eeq
Also in the gyrokinetic simulations, in accordance with Eq. (\ref{eqneue}), the equilibrium current density is assumed to be entirely due to the parallel electron velocity (we recall that, as discussed in Sec. \ref{ssec:connect}, the gyrokinetic admits, unlike the gyrofluid model, also a finite parallel ion flow).  
The tearing stability parameter \citep{Fur63} for the equilibrium is \cite{Por02}
\beq
\Delta'=  2 \frac{\left(5- k_y^2\right) \left( k_y^2+3\right)}{ k_y^2 ( k_y^2+4)^{1/2}}. 
\eeq
The equilibrium (\ref{equilgf}) is tearing unstable when $\Delta ' >0$, which corresponds to wave numbers $k_y = \pi m/L_y  < \sqrt{5}$.
In the development of the tearing process, after the saturation of the unstable dominant mode ($m=1$), eventually, a thinning  current sheet  located at the X point of the initial tearing configuration will form. In this evolving  current sheet, tearing-mode-like perturbations can develop to form secondary magnetic islands denoted as plasmoids, when they enter their non linear phase.

The fluid numerical solver SCOPE3D (Solver Collisionless Plasma Equations in 3D) (\cite{Gra2022pl}) is pseudo-spectral and the advancement in time is done through a third order Adams–Bashforth scheme. The numerical solver SCOPE3D  has been adapted to solve the gyrofluid equations. The gyrokinetic model is solved by {\tt AstroGK}~\citep{Num10}. Although {\tt AstroGK} employs some sophisticated techniques for the treatment of linear terms, it uses essentially the same pseudo-spectral and temporal schemes.

\section{Results on the plasmoid onset}

An extensive numerical simulation campaign, reported in the tables \ref{table1} and \ref{table2},  was carried out to study the physical conditions under which plasmoids appear.

\begin{table}
\begin{center}
\begin{tabular}{cccccccccc}
 & No.  & $\rs$ & $\Delta'$ & $\bee$ & $m_e/m_i$ & $\gamma_L$ & $\gamma_{max}$ &   Plasmoid \\
\hline
& $1_{GF1}$ & 0.3 & 14.3 & 0.2491  & 0.01 & 0.214 & 0.285 & One small  \\
& $1_{GF2}$ & 0.3 & 14.3 & 0.06228  & 0.0025 & 0.225 & 0.322 & No  \\
& $1_{F}$ & 0.3 & 14.3 & 0 & 0 & 0.230 & 0.337 & No  \\
& $2_{GF1}$ & 0.3 & 29.09 & 0.2491 & 0.01 & 0.211 & 0.342 & One plasmoid   \\
& $2_{GF2}$ & 0.3 & 29.09 & 0.1246 & 0.005 & 0.218 & 0.367 &  One plasmoid   \\
& $2_{GF3}$ & 0.3 & 29.09 & 0.06228 & 0.0025 & 0.231 & 0.378 &   One plasmoid  \\
& $2_{GF4}$ & 0.3 & 29.09 & 0.00622 & 0.0005 & 0.241 & 0.385 &    Several plasmoids    \\
& $2_{F}$ & 0.3 & 29.09 & 0 & 0 & 0.242 &  0.386 & Several plasmoids   \\
& $3_{GF1}$ & 0.5 & 14.3 & 0.692 & 0.01 & 0.286 & 0.334 & Several plasmoids \\
& $3_{GF2}$ & 0.5 & 14.3 & 0.3460 & 0.005 & 0.310 & 0.383 & Several plasmoids \\
& $3_{F}$ & 0.5 & 14.3 & 0 & 0 & 0.338 & 0.448 & Several plasmoids \\
& $4_{F}$ & 0.06 & 14.3 & 0 & 0 & 0.081 & 0.188 & No    \\
\end{tabular}
\end{center}
\caption{Gyrofluid and fluid simulations.}
\label{table1}
\end{table}
\begin{table}
\begin{center}
\begin{tabular}{cccccccccc}
 & No.  & $\rs$ & $\Delta'$ & $\bee$ & $m_e/m_i$ & $\gamma_L$ & $\gamma_{max}$ &   Plasmoid \\
\hline
& $1_{GK1}$ & 0.3 & 14.3 & 0.2491  & 0.01 & 0.2245 & 0.308 & One small  \\
& $1_{GK2}$ & 0.3 & 14.3 & 0.06228 & 0.0025 & 0.2438 & 0.342 & No   \\
& $2_{GK1}$ & 0.3 & 29.09 & 0.2491 & 0.01 & 0.2165 & 0.352 & One large   \\
& $2_{GK2}$ & 0.3 & 29.09 & 0.1246 & 0.005 & 0.2267 & 0.389 & One large \\
& $2_{GK3}$ & 0.3 & 29.09 & 0.06228 & 0.0025 & 0.2329 & 0.401 & One large \\
& $3_{GK1}$ & 0.5 & 14.3 & 0.692  & 0.01 & 0.3040 & 0.362 & One \\
& $3_{GK2}$ & 0.5 & 14.3 & 0.3460 & 0.005 & 0.3286 & 0.410 & One \\
& $3_{GK3}$ & 0.5 & 14.3 & 0.1730 & 0.0025 & 0.3472 & 0.453 & One  \\
& $4_{GK1}$ & 0.06 & 14.3 & 0.009965 & 0.01 & 0.08617 & 0.207 & No   \\
& $4_{GK2}$ & 0.06 & 14.3 & 0.002491 & 0.0025 & 0.08779 & 0.209 & No  \\
\end{tabular}
\end{center}
\caption{Gyrokinetic simulations.}
\label{table2}
\end{table}

Each simulation is identified by a code of the form $p_{F/GF/GK \, r}$, where $p$ and $r$ are integers and $F$, $GF$ and $GK$ indicate whether the simulation is carried out in the fluid limit, with the gyrofluid model or with gyrokinetic model, respectively. 
For all the simulations, the value of the electron skin depth is fixed to $d_e=0.085$. Simulations with the same number $p$ are characterized by the same values of $d_e$, $\rho_s$ and $\Delta'$. For a fixed $p$, different values of the index $r$, on the other hand, indicate different values of $\beta_e$ (and, consequently, of $m_e/m_i$), with $\beta_e$ decreasing as $r$ increases. Not all the simulations of  table \ref{table1} have a corresponding simulation in table \ref{table2} and viceversa, although this is the case for most of the simulations. In particular, we point out that, because gyrokinetic simulations always have a finite value of $\beta_e$, strictly speaking there is no gyrokinetic counterpart for the fluid simulations, which formally correspond to the $\beta_e \rightarrow 0$ limit.

For all the gyrokinetic simulations, the temperature ratio is set to $\tau = 10^{-3}$, where the ion Larmor radius is  $\sqrt{\tau}\rho_s$. As mentioned before, the gyrofluid model assumes $\tau \rightarrow 0$. Therefore, in both the gyrofluid and gyrokinetic approach, the ion Larmor radius effects are neglected. 

As a first general comment, we observe, by comparing gyrofluid and gyrokinetic simulations with the same indices $p$ and $r$, that, in terms just of appearance or absence of plasmoids, gyrofluid simulations agree with the gyrokinetic ones. Therefore, in this respect, we can conclude that the quasi-static closure for the electrons and the suppression of ion gyrocenter fluctuations, do not affect critically the stability of the nonlinear current sheet. However, as will be discussed in the next Sections, differences appear in terms of the number and size of plasmoids. In particular, when more than one plasmoid is observed, this is indicated in table \ref{table1}, generically, as 'several plasmoids'. The number of plasmoids in the same simulation can indeed vary in time, as plasmoids can form at different times and pairs of plasmoids can merge into a single one.

The series of simulations for $p=1$ and $p=2$ will be used to show the effect of $\beta_e$ on the formation of plasmoids, while the simulations $p=3$ and $p=4$ will be used to verify that the regime $\rho_s \ll d_e$ promotes the formation of plasmoids.

\subsection{Growth rates}

Before discussing in detail the plasmoid instability, we briefly comment about the linear growth rate of the tearing mode excited by the perturbation of the initial equilibria (\ref{equilgf}). 
In the tables, we reported the value of the linear and maximum growth rate of the tearing instability, evaluated measuring the following quantity at the X-point
\beq 
\gamma = \frac{d}{dt} \log\left| \apar^{(1)} \left(\frac{\pi}{2},0,t \right)  \right|. \label{numericalgrowthrate}
\eeq
The two approaches give close growth rate values for small $\beta_e$. One can note a discrepancy on the value of $\gamma_{max}$ when $\beta_e $ is large. This suggests that, for large values of $\beta_e$ and during the non-linear phase, the efficiency of the gyrofluid model to reproduce the gyrokinetic results becomes limited. As commented in Sec. 2.3, one reason for this might be the absence of ion gyrocenter density and parallel velocity fluctuations, which occurs in the gyrofluid model, even for large $\beta_e$, due to the imposed condition (\ref{ioncond}). \\
\begin{figure}
    \centering
\includegraphics[trim=50 0 0 0, scale=0.6]{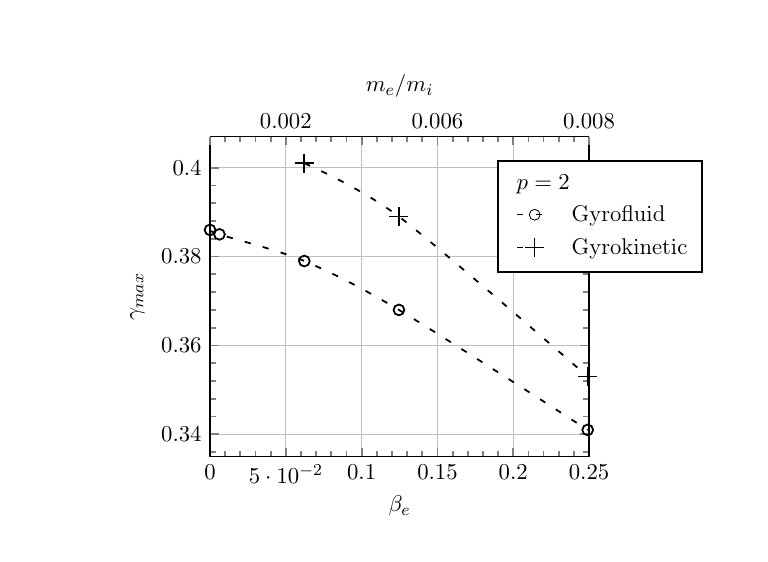}\hspace{-0.3cm}
\includegraphics[trim=50 0 0 0, scale=0.6]{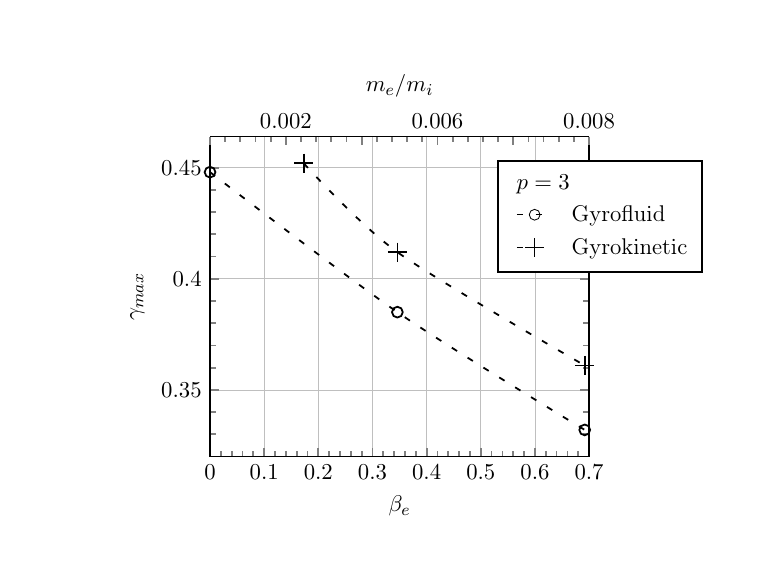}
\caption{Maximum growth rates of the collisionless tearing mode as a function of $\beta_e$,  for the cases $p=2$ and $p=3.$}
    \label{fig:gr}
\end{figure}
As shown on Fig. \ref{fig:gr}, the gyrofluid and gyrokinetic simulations yield the same dependence of the growth rates on the parameters, with gyrofluid simulations slightly underestimating the growth rate, in general.
By comparing the growth rate results, for a fixed mass ratios, of simulations $p=1-3-4$, we note that increasing $\beta_e$ and $\rho_s$, as $\rho_s \sim \sqrt{\beta_e/2}$, destabilizes the tearing mode. Increasing these parameters can be seen as fixing the background density, the ion mass and the guide field amplitude, while increasing the electron temperature. It was shown numerically in \cite{Num15, Gra22flr} that, in this latter situation, the linear tearing growth rate is first ruled by the destabilizing effect of the sonic Larmor radius. However, in cases where the electron temperature is high enough for the effects of $\rho_e$ to take over those of $\rho_s$, the linear growth rate is damped. Here, we find ourselves in the first case, for which the effects of the sonic Larmor radius are visibly dominant.

\subsection{Remarks on the numerical resolution}
\label{sec:resolution}

It is important to anticipate the role of the resolution in this study. In the forming nonlinear current sheet, tearing modes grow and can become unstable at different times. The current sheet can therefore be broken by multiple dominant modes, and the number of plasmoids is highly sensitive to the resolution used.  Given that the fluid simulations $2_F$ and $3_F$ were those which allowed the formation of several plasmoids, we carried out resolution tests  with the gyrofluid code on these two simulations to determine the necessary number of points along $y$, that does not prevent the growth of large mode numbers.

Table \ref{table3} reports the number of visible plasmoids for simulation $2_F$ as a function of the number of points and indicates their order of appearance. The convergence is reached for a resolution of $2304^2$. \\

\hfill\\
\begin{table}
\begin{tabular}{cccc}
 & $n_x \times n_y$  & \# plasmoids & Comments on the order of appearance\\
\hline
& $200 \times 120$ & 1 & 1 at the center \\
& $200^2$ & 3 & 2 symmetrically with respect to the center then 1 at the center  \\
& $200 \times 400$ & 3 & 2 symmetrically with respect to the center then 1 at the center \\
& $2304^2$ & 7 & 6 symmetrically with respect to the center then 1 at the center  \\
& $3400\times4800$ & 7 & 6 symmetrically with respect to the center then 1 at the center  \\
\end{tabular}
\caption{number of visible plasmoids for simulation $2_F$ for different grids.}
\label{table3}
\end{table}
\hfill\\

For $3_F$, which is close to marginal stability, a spatial discretization smaller than $2 L_y / n_y \sim 0.0078$ was needed. Unfortunately, it is not foreseeable to perform gyrokinetic simulations with such a high resolution. Therefore, a grid of $256 \times 128$ points has been used for all the gyrokinetic runs. We compared these gyrokinetic simulations to fluid$/$gyrofluid simulations performed with a nearly identical resolution. However, since the fluid code is much less demanding in computation time, we also performed the fluid simulations with grids up to $2304^2$ points.

\subsection{Effect of $\beta_e$ on the plasmoid onset}  \label{ssec:betae}

In this Section we present how the $\beta_e$ parameter changes the characteristics of the forming current sheet and promotes the plasmoid formation.  We measure the current sheet aspect ratio using the current density $j_{\parallel}$. The length is defined such that the current distribution from the X point to $L_{\rm{cs}}/2$ equals a specific value $\alpha$
\beq \label{stdev}
\frac{1}{N} \sum^N_{i=1} \left(j_{\parallel}|_X - j_{\parallel}(0, i \Delta y ,t)\right)^2 = \alpha j_{\parallel}|_X,
\eeq 
where $\Delta y$ is the mesh length along $y$ and $N$ indicates the number of points from $y=0$ to $y=L_{\rm{cs}}/2$. The constant $\alpha$ is taken to $1/3$ as it gives good measurement of the length of the region with a strong current. Formula (\ref{stdev}) allows to apply a single consistent method for all the simulations, while taking into account the reduction of the current intensity along the layer.  The width of the current sheet corresponds to the distance between the two points along $x$ where the value of $j_{\parallel}$ takes the same value as at the point $(0,L_{\rm{cs}}/2)$.\\

\begin{figure}
    \centering
\hspace{-1.2cm}\includegraphics[trim=0 0 0 0, scale=0.21]{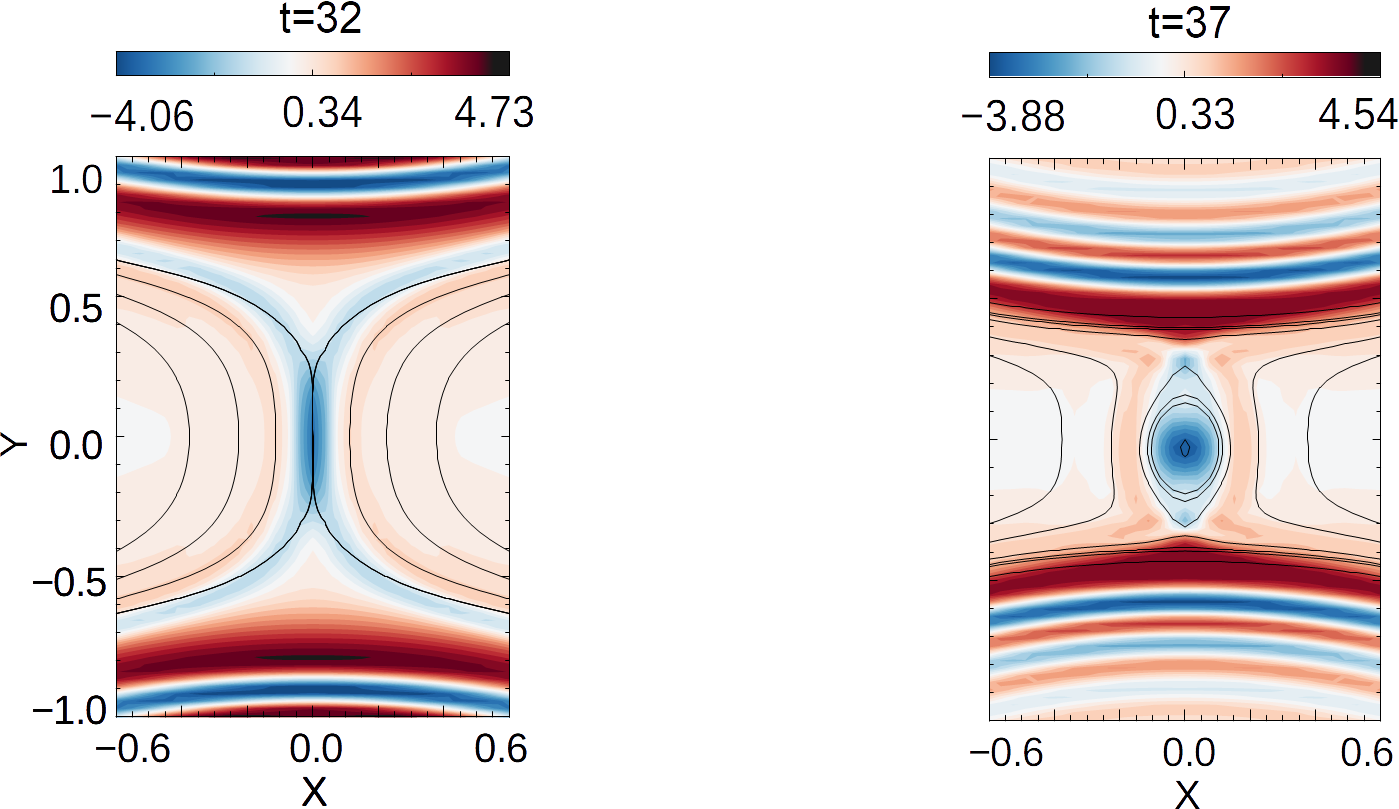}\\
\includegraphics[trim=0 0 0 0, scale=0.24]{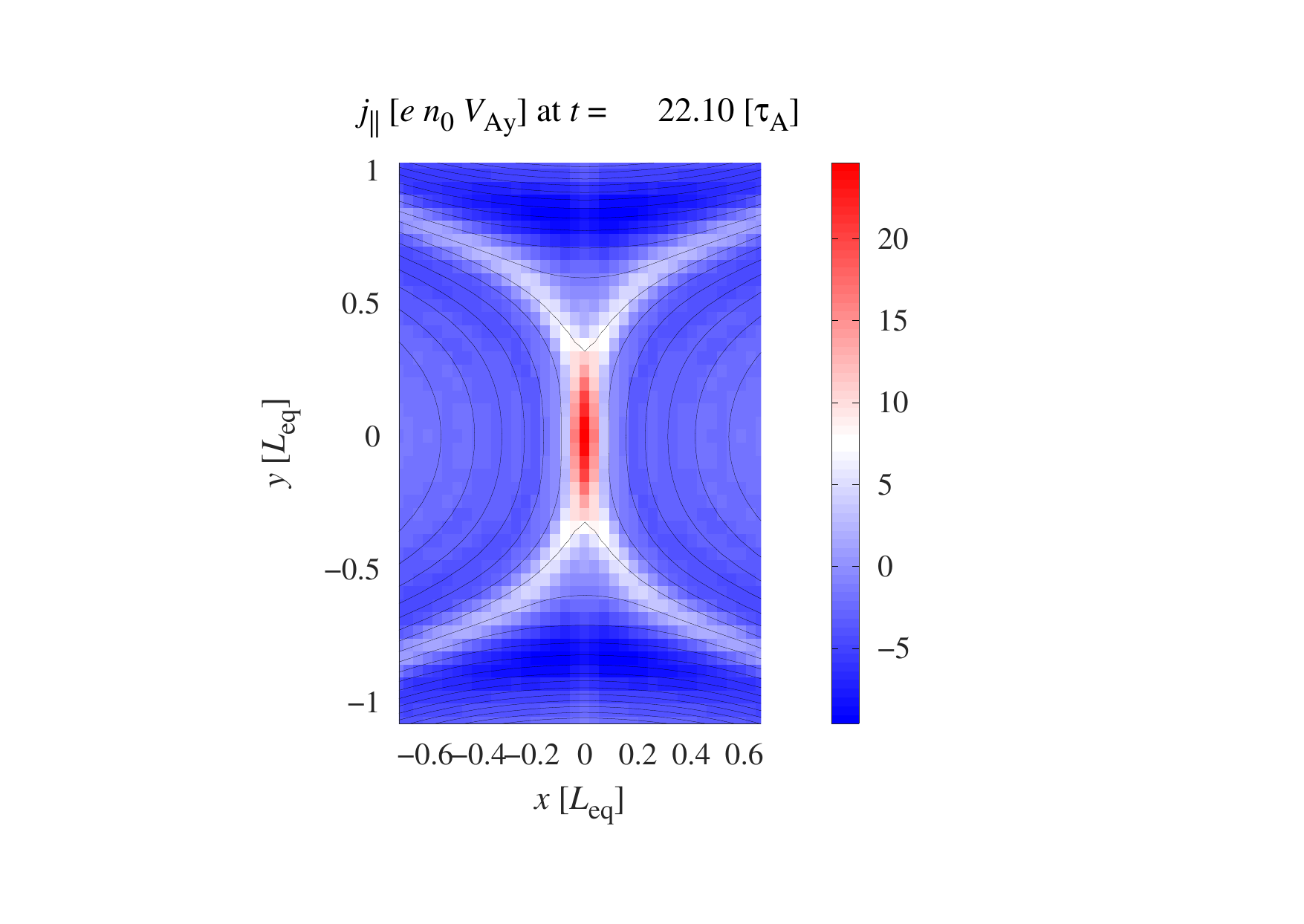}\hspace{-2.3cm}
\includegraphics[trim=0 0 0 0, scale=0.24]{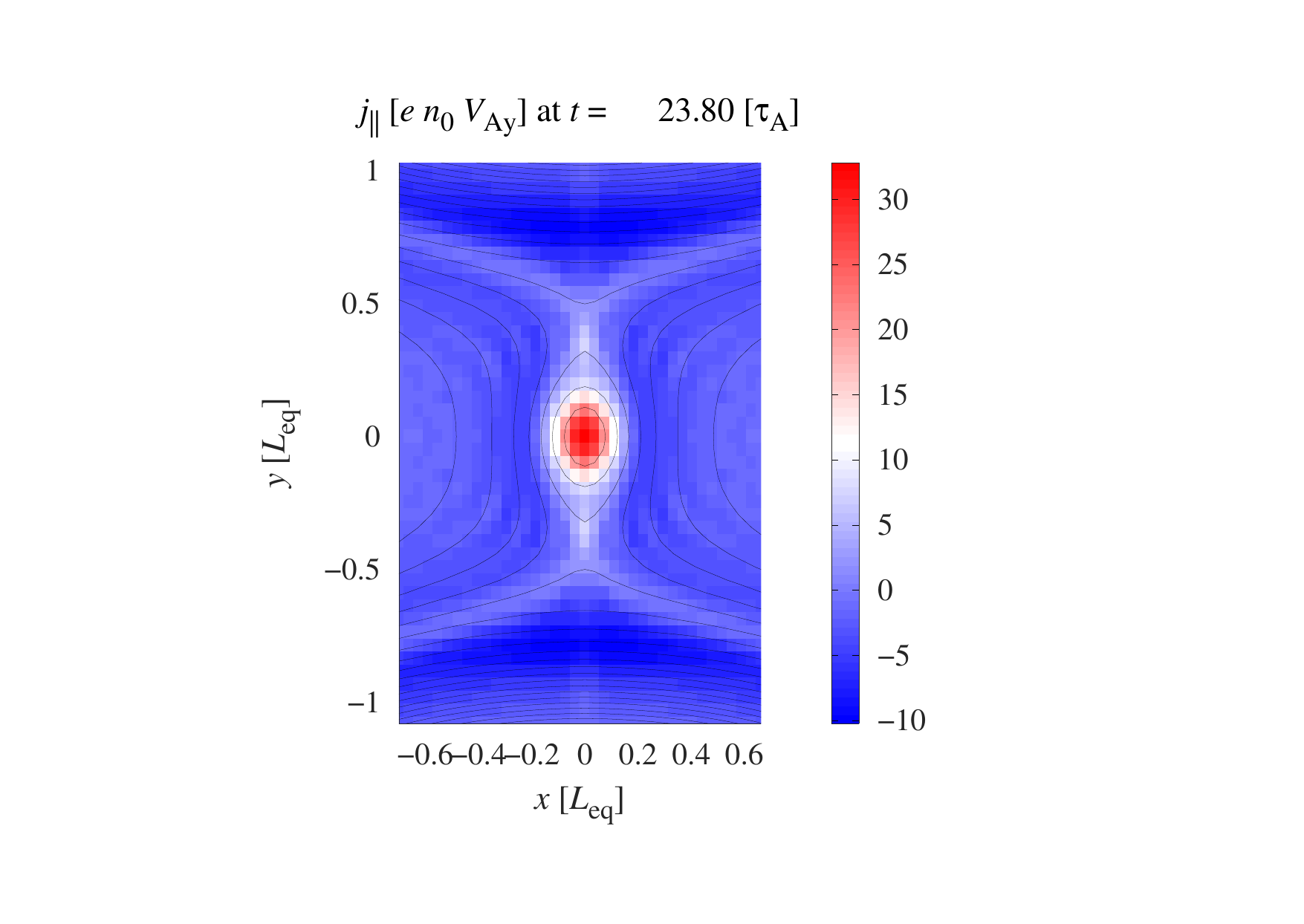}
\caption{Top: Contour of the parallel electron velocity $u_e$ (proportional to the parallel current density) for simulation $1_{GF1}$. Bottom:  Contour of the parallel current density $j_{\parallel}$ of simulation $1_{GK1}$. Isolines of the magnetic potential are superimposed on all the contours.   }
    \label{fig:contcase1.3}
\end{figure}

We focus first on the comparison of the series of simulations for $p=1$, starting with the higher $\beta_e$ case, for which $\bee=0.2491$.
The contour plots of the parallel electron velocity $u_e$ (proportional to the parallel current density), for the gyrofluid simulation $1_{GF1}$ and of the current density, $j_{\parallel}$, for the gyrokinetic simulations $1_{GK1}$, are shown on Fig. \ref{fig:contcase1.3}. Isolines of the magnetic potential, showing the topology of the magnetic field, are overplotted.
Both approaches indicate the formation of a plasmoid. For the fluid simulation, the aspect ratio is  $A_{\rm{cs_f}}=4.90$. In the gyrokinetic case, we measure $A_{\rm{cs_k}}=4.03$. There is a constant difference between the value of the gyrofluid and gyrokinetic aspect ratio, which is explained by the difference in resolution. Their evolution according to the parameters, on the other hand, are in agreement. \\ 
\begin{figure}
    \centering
\hspace{-1.1cm}\includegraphics[trim=0 0 0 0, scale=0.21]{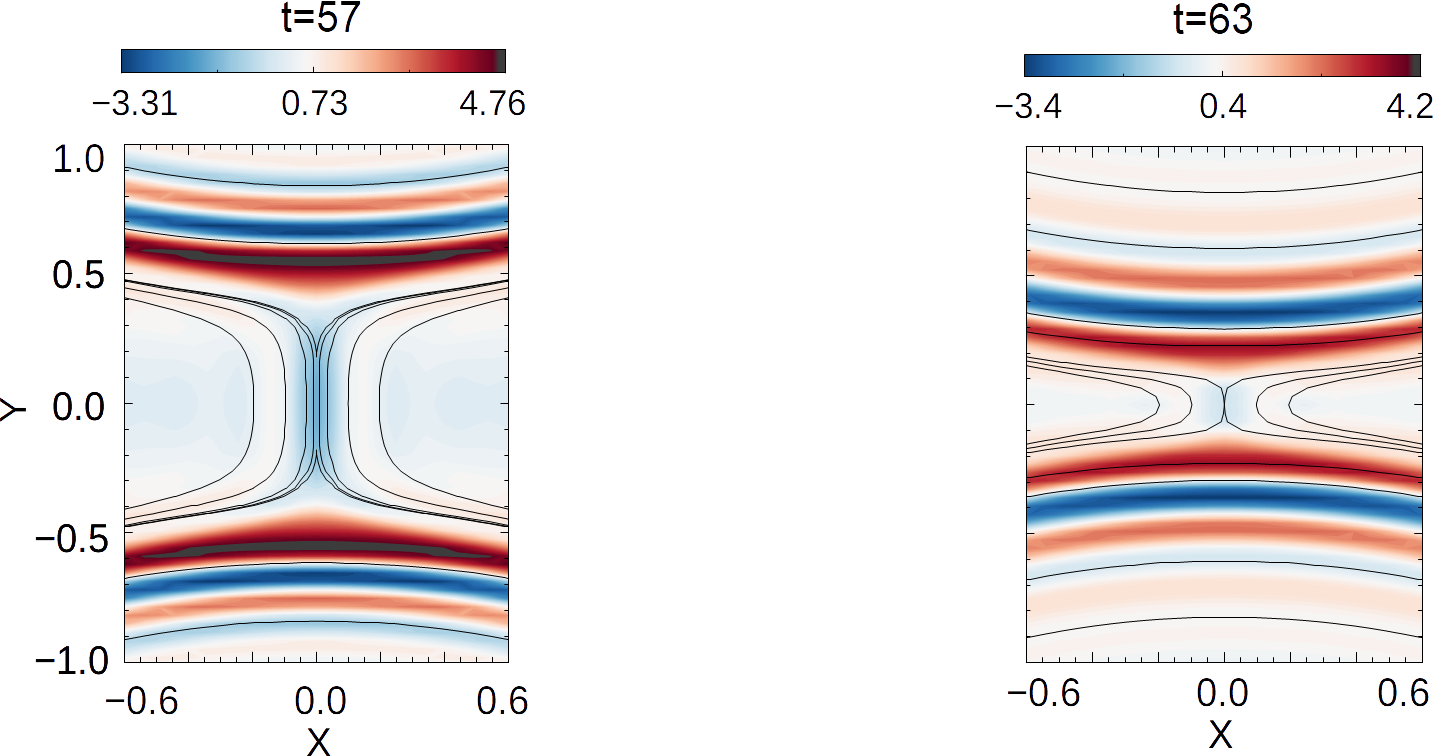}\\
\includegraphics[trim=0 0 0 0, scale=0.24]{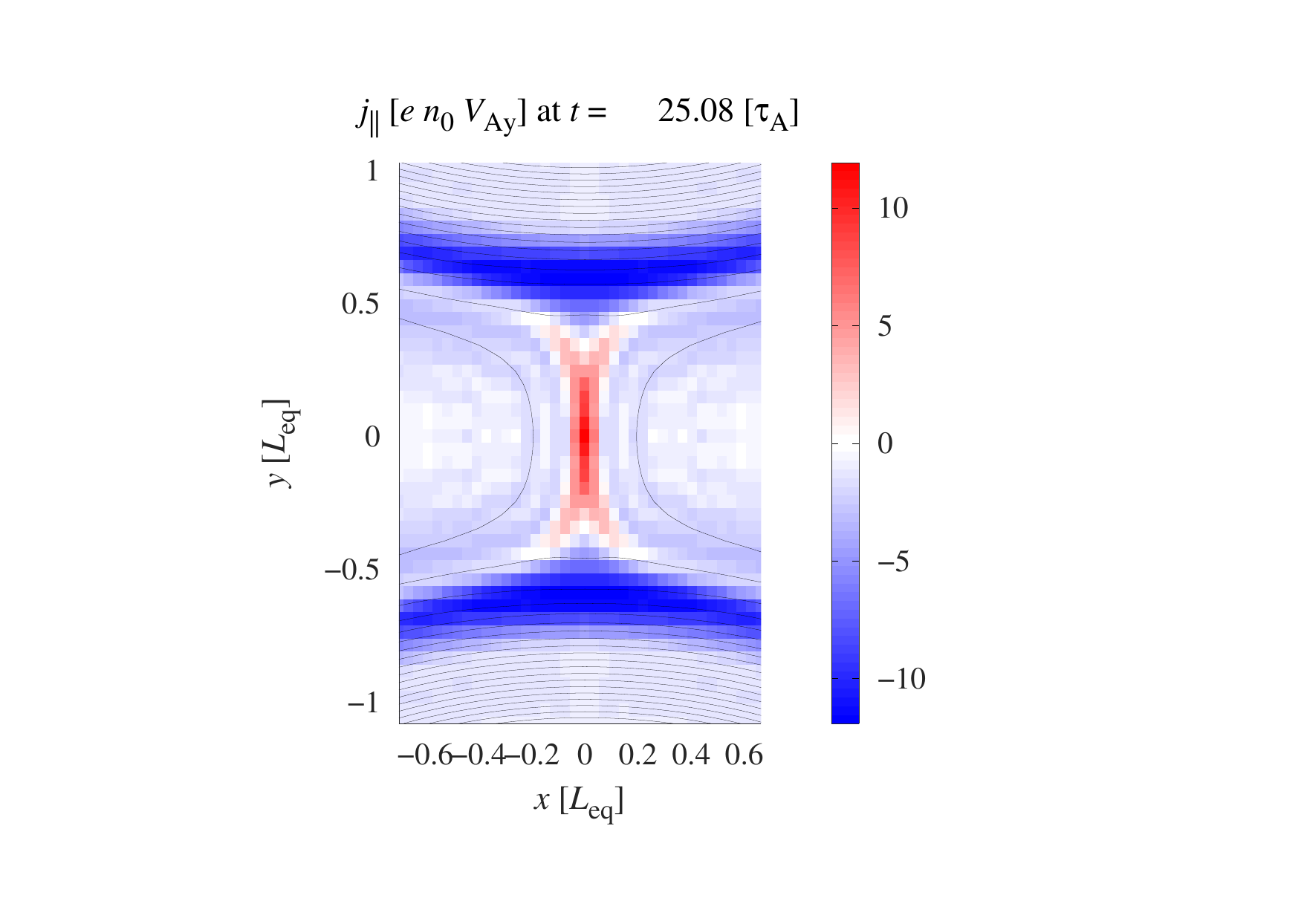}\hspace{-2.3cm}
\includegraphics[trim=0 0 0 0, scale=0.24]{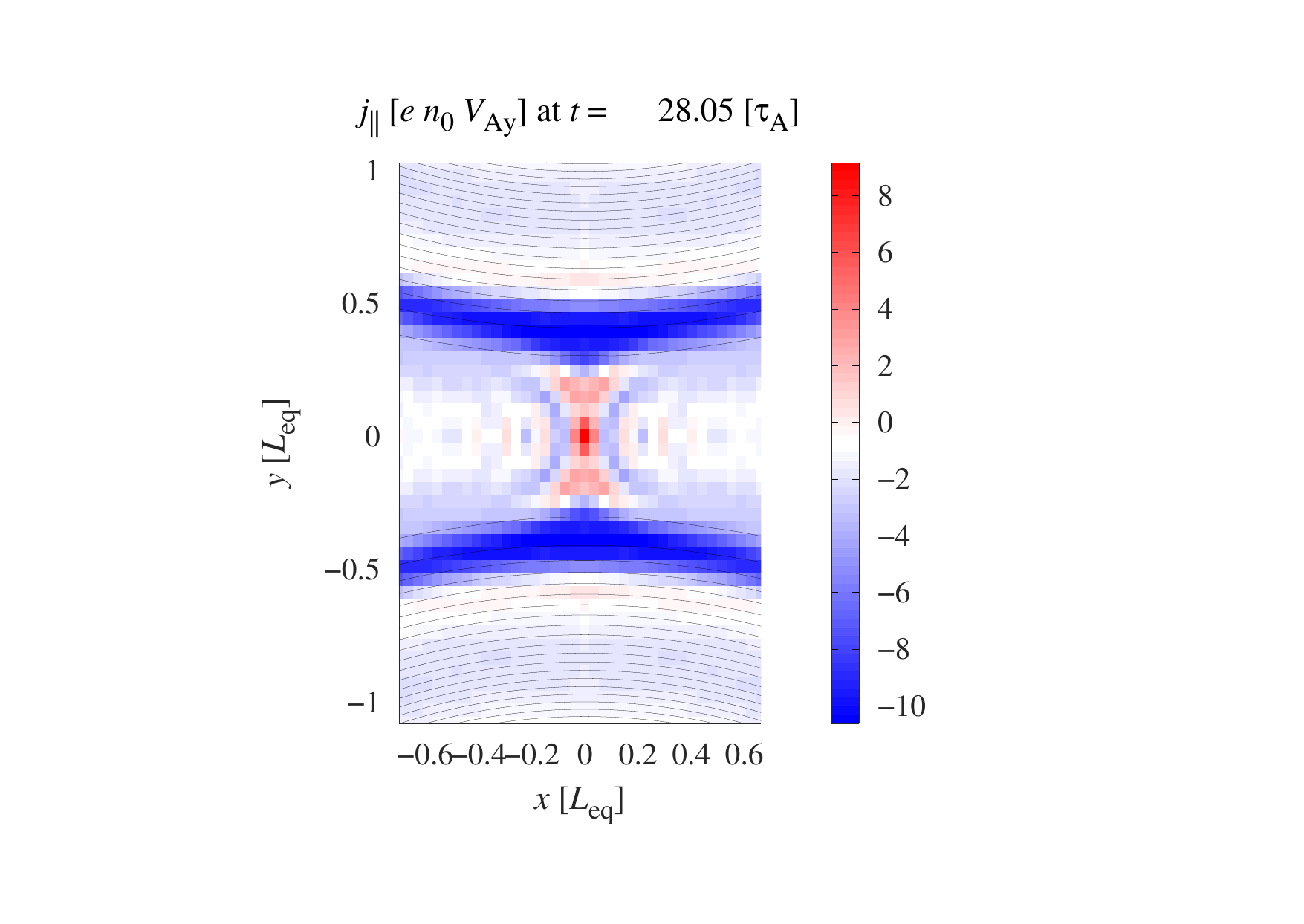}
 \caption{Top: Contour of the electron velocity $u_e$ (proportional to the parallel current density) for simulation  $1_{GF2}$. Bottom: Contour of the parallel current density $j_{\parallel}$ of simulation $1_{GK2}$.  Isolines of the magnetic potential are superimposed on all the contours.  }
    \label{fig:contcase1.1} 
\end{figure}
For the lowest $\beta_e$ cases, for which $\bee=0.06228$, the contour plots of the simulations $1_{GF2}$ and $1_{GK2}$, are shown on Fig. \ref{fig:contcase1.1}. The two simulations lead to the formation of a stable current sheet having an aspect ratio decreasing in time. The maximum aspect ratio is reached when the growth rate has reached its maximum value and the process enters the saturation phase. From the gyrofluid simulation we measured a maximum aspect ratio $A_{\rm{cs_f}}=5.11$. In the gyrokinetic case, the aspect ratio is $A_{\rm{cs_k}}=4.14$.
The measured aspect ratio are very close to those obtained for $\beta_e=0.2491$, and yet, no plasmoids develop.  In this first series of tests we are at the frontier between stability and instability, and the role of $\beta_e$ seems crucial to switch to an unstable case. \\

\begin{figure}
\includegraphics[trim=20 0 0 0, scale=0.81]{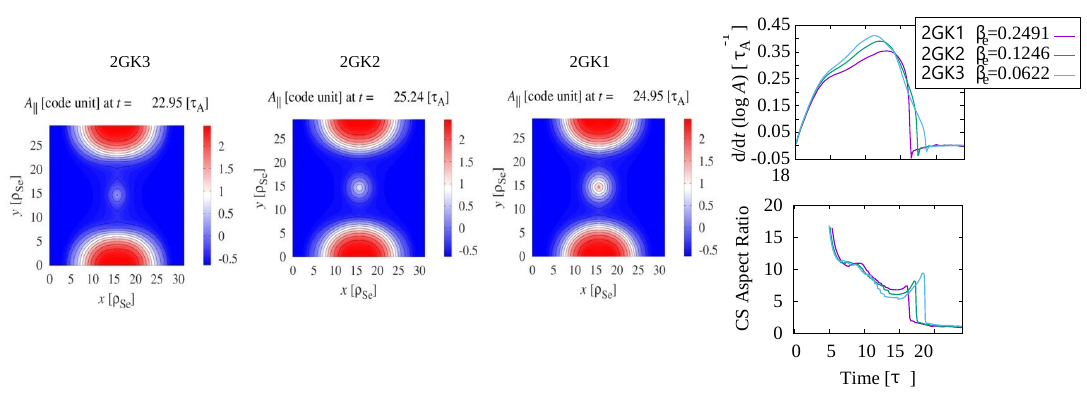}
 \caption{Left: contours of $\apar$ for simulations $2_{GK3}$, $2_{GK2}$ and $2_{GK1}$. Right: growth rate evolution as function of time and aspect ratio of the forming  current sheet  as a function of time for the same simulations. }
    \label{fig:case2gk}
\end{figure}

In the series of simulations for $p=2$, the idea is to consider the same parameters as those for $p=1$ but with a longer forming current sheet.
Since highly unstable primary reconnecting modes favour the  formation of extended secondary current sheets we consider  a larger domain size along the y direction, with $L_y=1.4 \pi$, that corresponds to $\Delta'=29.9$.
The other parameters are kept the same. 
In this case, even the small/negligible $\beta_e$ simulations become plasmoid unstable. Figure \ref{fig:case2gk} shows the plasmoids and the growth rate evolutions obtained for simulations $2_{GK1}$ - $2_{GK3}$.  On Fig. \ref{fig:case2gk}, in order to better show the plasmoid size, we show  the contour plots of the magnetic potential $\apar$.  The magnetic potential contour is shown as the plasmoid reaches its maximum size, which occurs in the saturation phase of the tearing instability. 
It can be seen that increasing $\beta_e$ results in larger sized plasmoids, although, from the aspect ratio measurement, increasing $\beta_e$ reduces the aspect ratio obtained just before the plasmoid onset.
Here, increasing $\beta_e$ (considering therefore larger mass ratios)  seems to have a similar effect to increasing $\rho_s$, and allows the plasmoid instability to grow better in current sheets whose dimensions are not particularly favourable (low aspect ratio).  A recent result obtained in a 2D, collisionless, fluid model (\cite{Graso22}) has shown that rhos significantly enlarge the spectrum of the linear unstable reconnecting modes that develop in presence of a sheared flow  and magnetic field

\begin{figure}
    \centering 
\includegraphics[trim=0 0 0 0, scale=1.1]{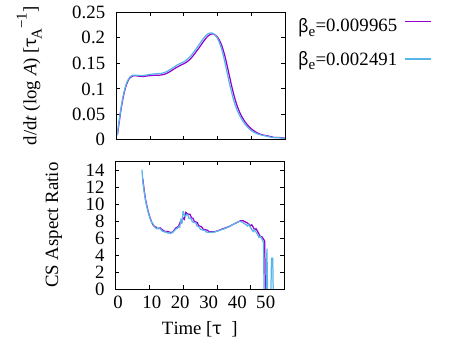}
 \caption{Growth rate evolution as function of time (top) and aspect ratio of the forming current sheet as a function of time (bottom) for the simulations $4_{GK1}$ and $4_{GK2}$.}
 \label{aspgk}
\end{figure}

In comparison, Fig. \ref{aspgk} shows the aspect ratio and the growth rate obtained for the simulations  $4_{GK1}$ and $4_{GK2}$. For this set of simulations, the effects of $\beta_e$ are negligible and the parameters $\rho_s$ and $\Delta'$ are smaller than those of simulations $2_{GK1}$ - $2_{GK3}$. Nevertheless, despite a very different set of parameters, the two set of simulations lead to the formation of current sheets whose aspect ratio is almost identical. Yet, unlike cases $2$, cases $4$ remain stable. 
If we assume, as done in the theory by \cite{Com16} and \cite{Gra2022pl}, that the growth of the perturbation in the secondary current sheet does not depend on the initial tearing equilibrium (and that therefore the initial $\Delta'$ has no role in the plasmoid size) then plasmoid formation is determined only by the values of $\rho_s$ and $\beta_e$ which are greater in case $2$.

\begin{figure}
    \centering
\includegraphics[trim=0 0 0 0, scale=0.2]{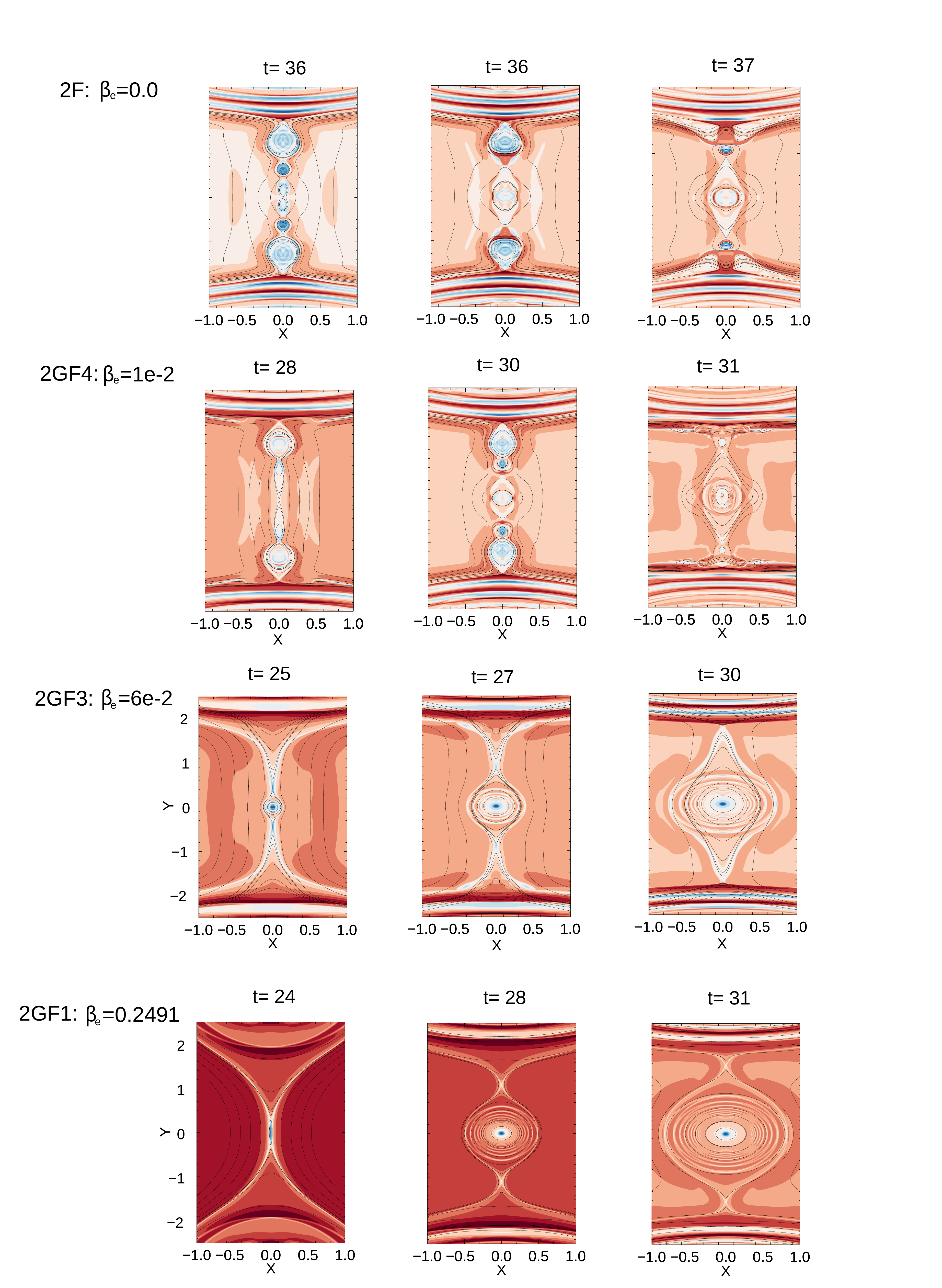}
\caption{Contour of $u_e$ with isolines of $\apar$. From top to bottom panels:
:  $2_{F}$ ($2304^2$), $2_{GF4}$, $2_{GF3}$, $2_{GF1}$ ($2304\times2400$).}
\label{fig:case2fl}
\end{figure}

Figure \ref{fig:case2fl} shows in detail the evolution of the instability for 
$2_{F}$, $2_{GF4}$, $2_{GF3}$ and $2_{GF1}$ having the highest resolution. For negligible $\beta_e$ we can see several plasmoids forming in a row, whereas, for  $\beta_e>0.06$ we see only a single central plasmoid.  The effects of $\beta_e$ eventually prevent the development of large modes inside the current sheets.

\subsection{Validation of the plasmoid regime for $\rho_s \gg d_e$}   \label{ssec:fluid}

A theory and numerical study developed by \cite{Gra2022pl} stated that, for a current sheet close to marginal stability, the regime $\rho_s \gg d_e$ promotes the plasmoid formation. In this Ref., the simulations were carried out with the fluid model (\ref{fluid1}) - (\ref{fluid2}) which assumes a negligible mass ratio and a negligible $\beta_e$. In this subsection, we present a gyrokinetic validation of these results. In addition to observing a possible role played by the closure, we also compare the fluid results with those including a finite mass ratio of $m_e/m_i=0.005$, and consequently a small $\beta_e$. Moreover, as already recalled, the evolution of ion quantities such as $N_i$ and $U_i$, prevented by the gyrofluid model, but present in the gyrokinetic simulations, might in principle also play a role. 
Therefore, in this subsection we focus on the low $\beta_e$ regime and compare the simulation set for $p=3$, for which $\rho_s \gg d_e$, with the simulation set for $p=4$, for which $\rho_s \ll d_e$. These two sets of simulation lead to the formation of a secondary current layer close to the instability threshold.

\begin{figure}
    \centering
\hspace{-1.cm}\includegraphics[trim=0 0 0 0, scale=0.2]{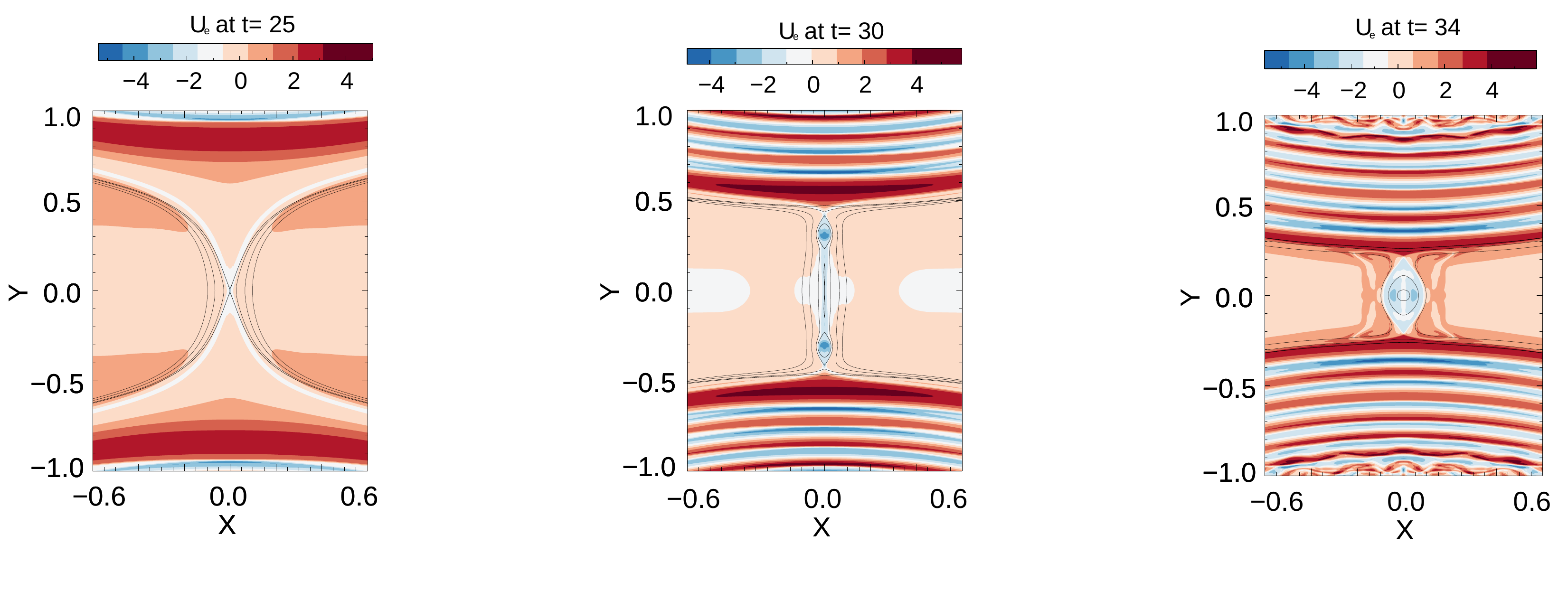}\\
\includegraphics[trim=100 0 0 0, scale=0.22]{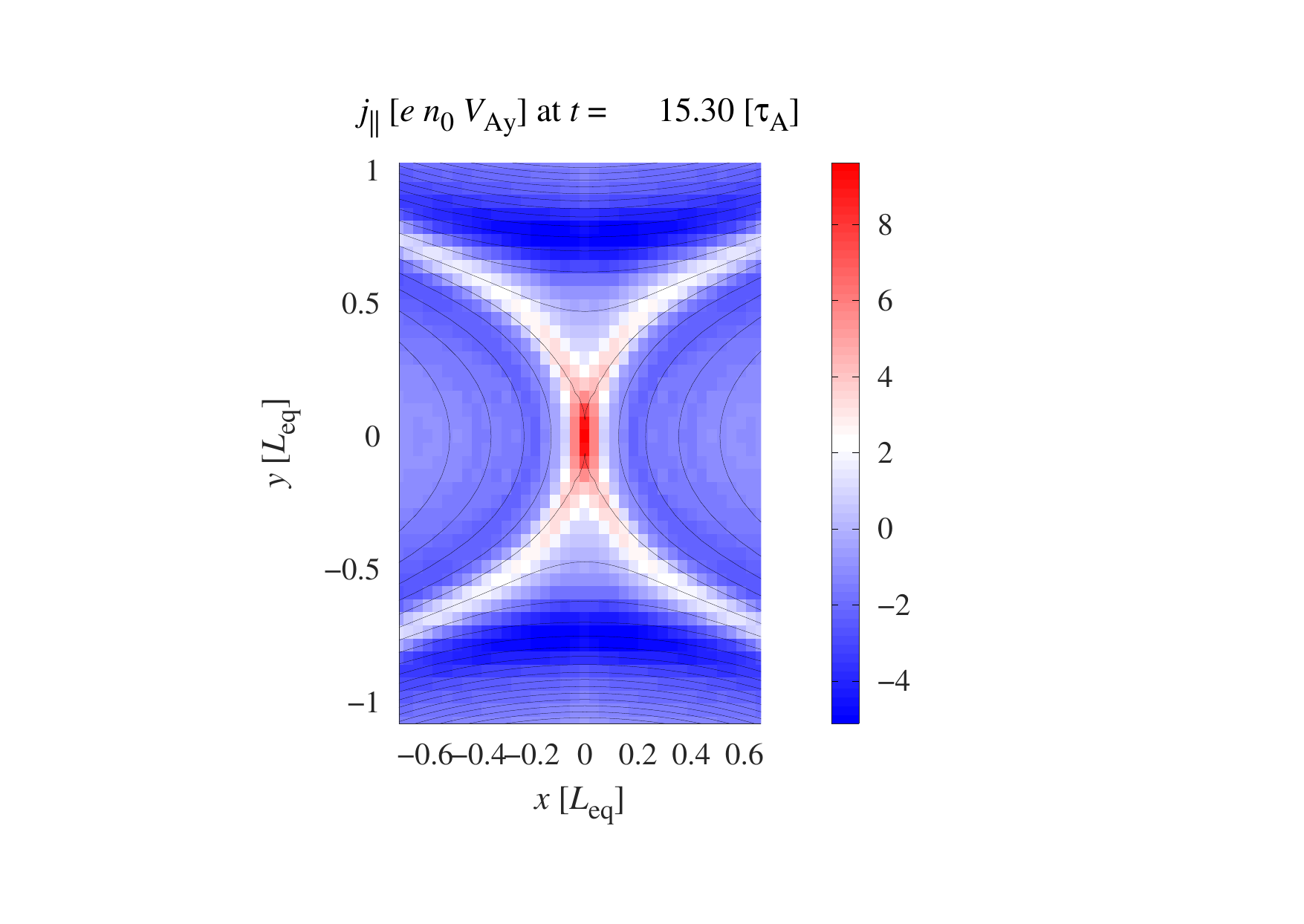} \hspace{-2.5cm}\includegraphics[trim=0 0 0 0, scale=0.22]{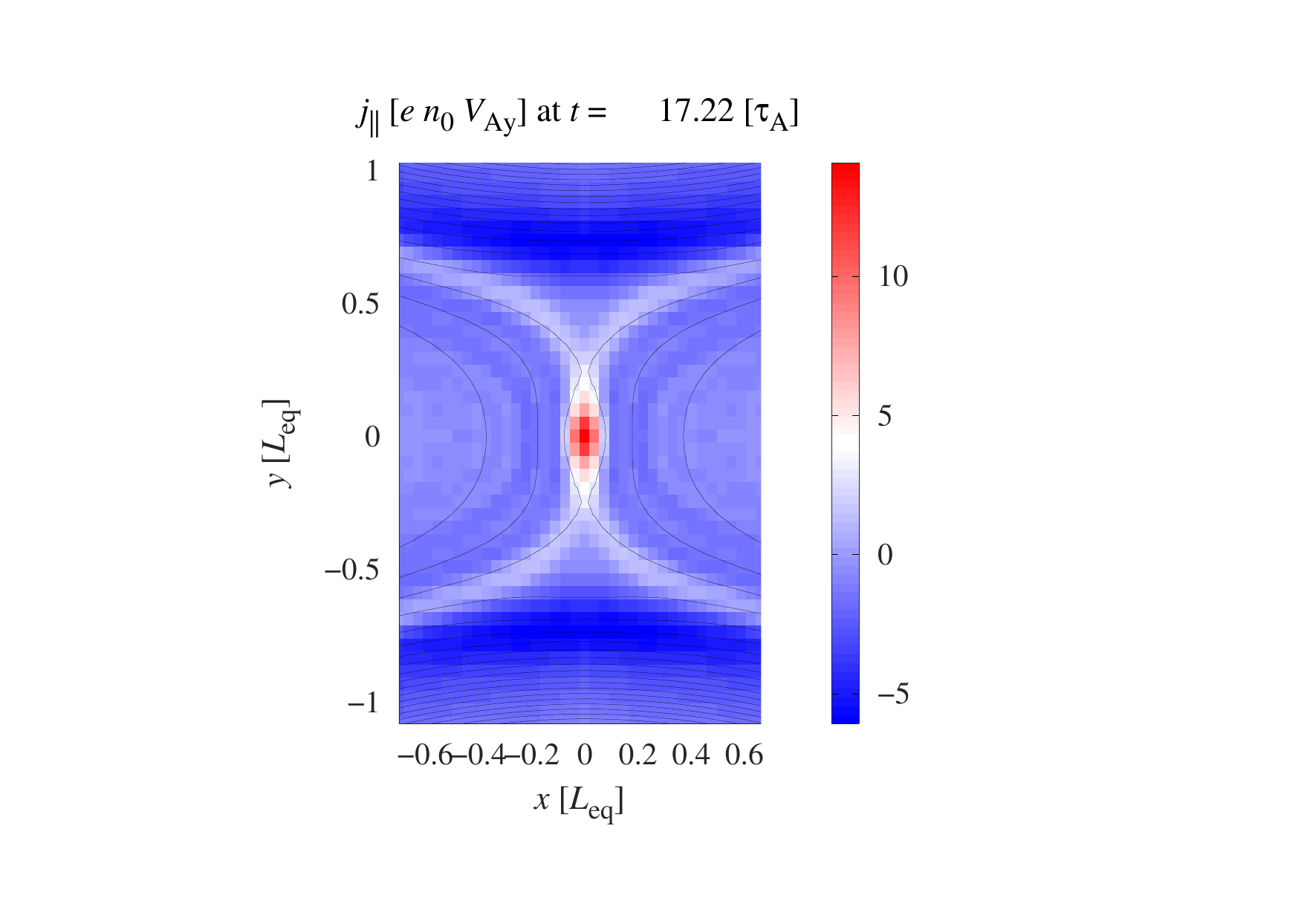} \hspace{-2.5cm}\includegraphics[trim=0 0 100 0, scale=0.22]{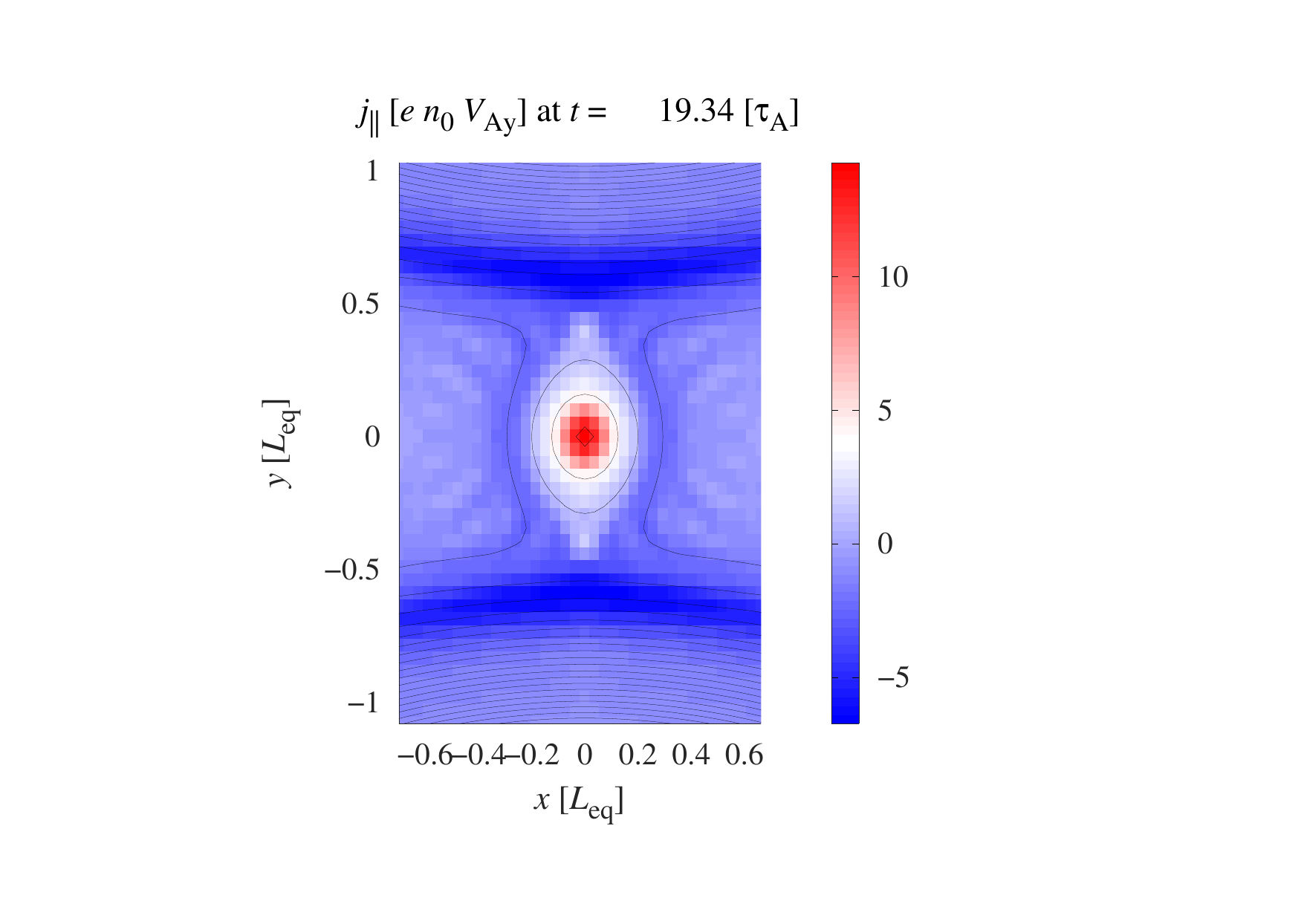}
  \caption{Top: contour of the parallel current density $u_e=U_e$ for simulation $3_{F}$ ($1728^2$). Bottom: Contour of the parallel current density $j_{\parallel}$ of simulation $3_{GK3}$.   Isolines of the magnetic potential are superimposed on all the color maps. }
    \label{fig:case3contour}
\end{figure}

Figure \ref{fig:case3contour} shows the evolution of the instability for the simulations $3_{GK3}$ (lowest $\beta_e$ gyrokinetic case of this series) and $3_F$. For the two approaches, the current sheet becomes plasmoid unstable. 
Also in this case the resolution plays an important role. With a resolution of $1728^2$, three plasmoids were visible in the simulations $3_F$. However, the same fluid simulation performed with a resolution $500\times360$ shows only one plasmoid. Since a resolution higher than that was not foreseeable with the gyrokinetic code, we used a grid of $256\times128$ points that allowed to observe one single plasmoid at the center. 
In the regime $\rho_s \gg d_e$, the current aligns with the magnetic field lines, thus forming a cross shaped current sheet \citep{Caf98}. This behavior is retrieved by the gyrokinetic simulation. 

\begin{figure}
    \centering
\hspace{-1.cm}\includegraphics[trim=0 0 0 0, scale=0.22]{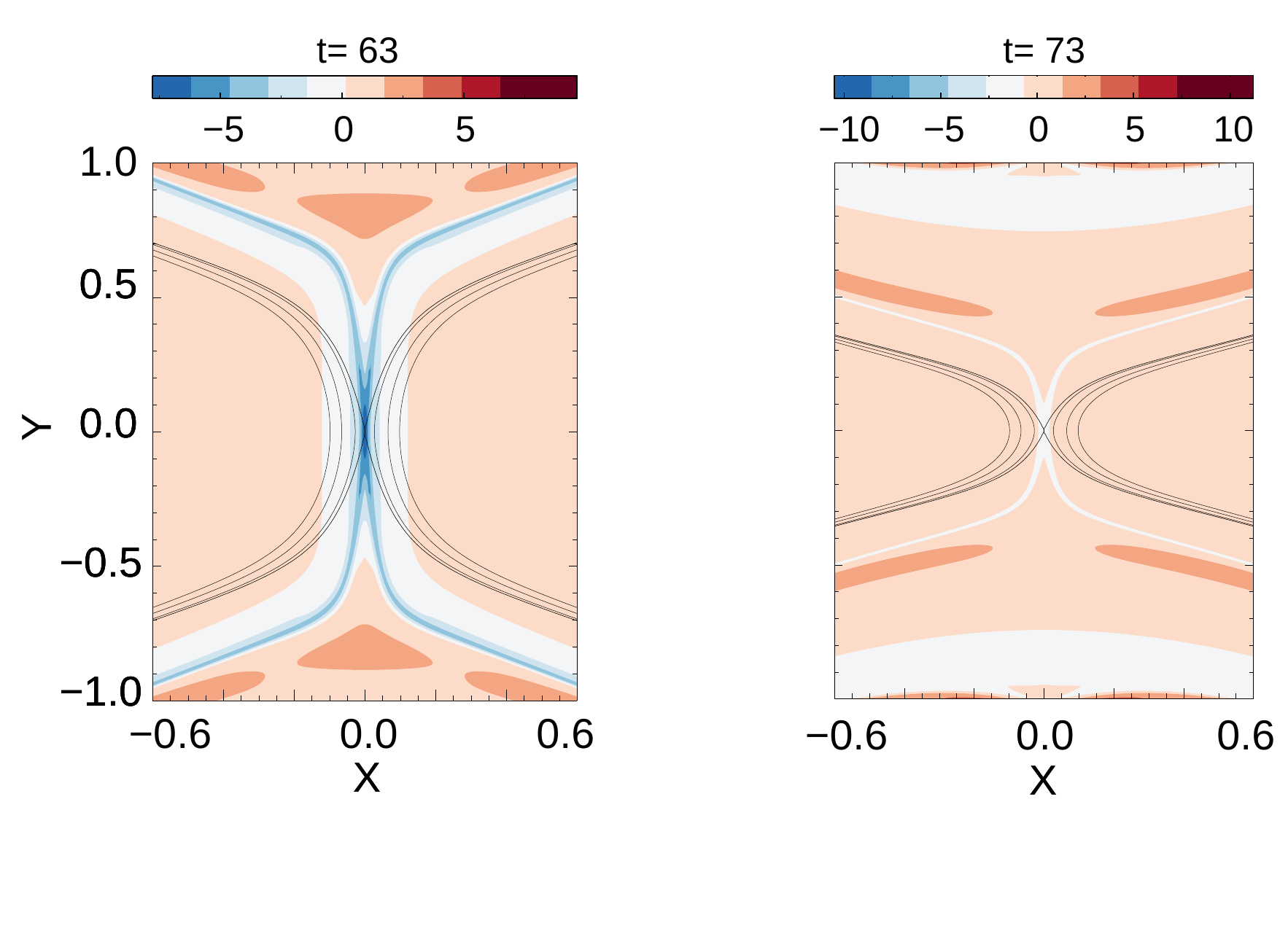}   \\   
\includegraphics[trim=0 0 0 0, scale=0.24]{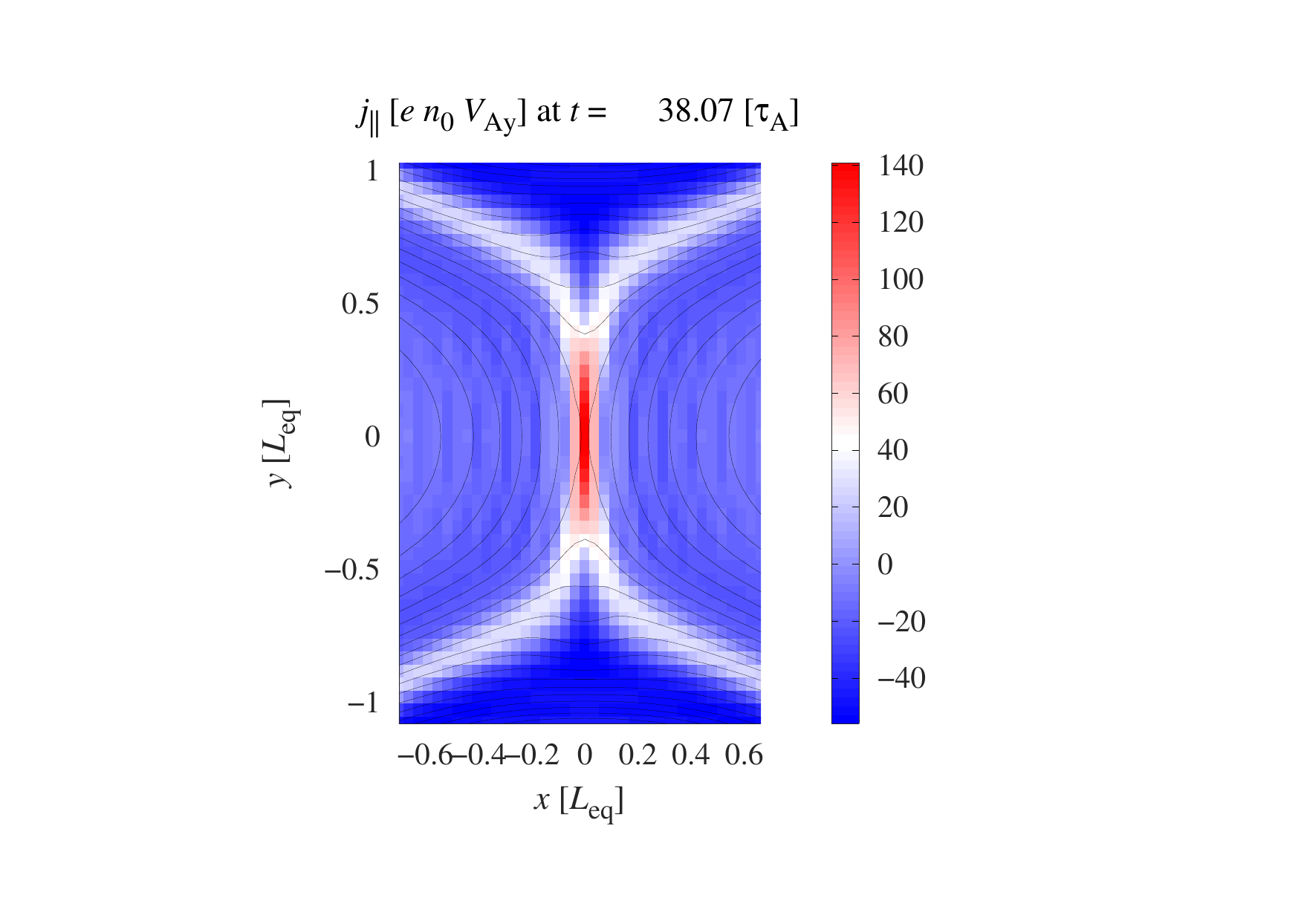}\hspace{-3.2cm}
\includegraphics[trim=0 0 0 0, scale=0.24]{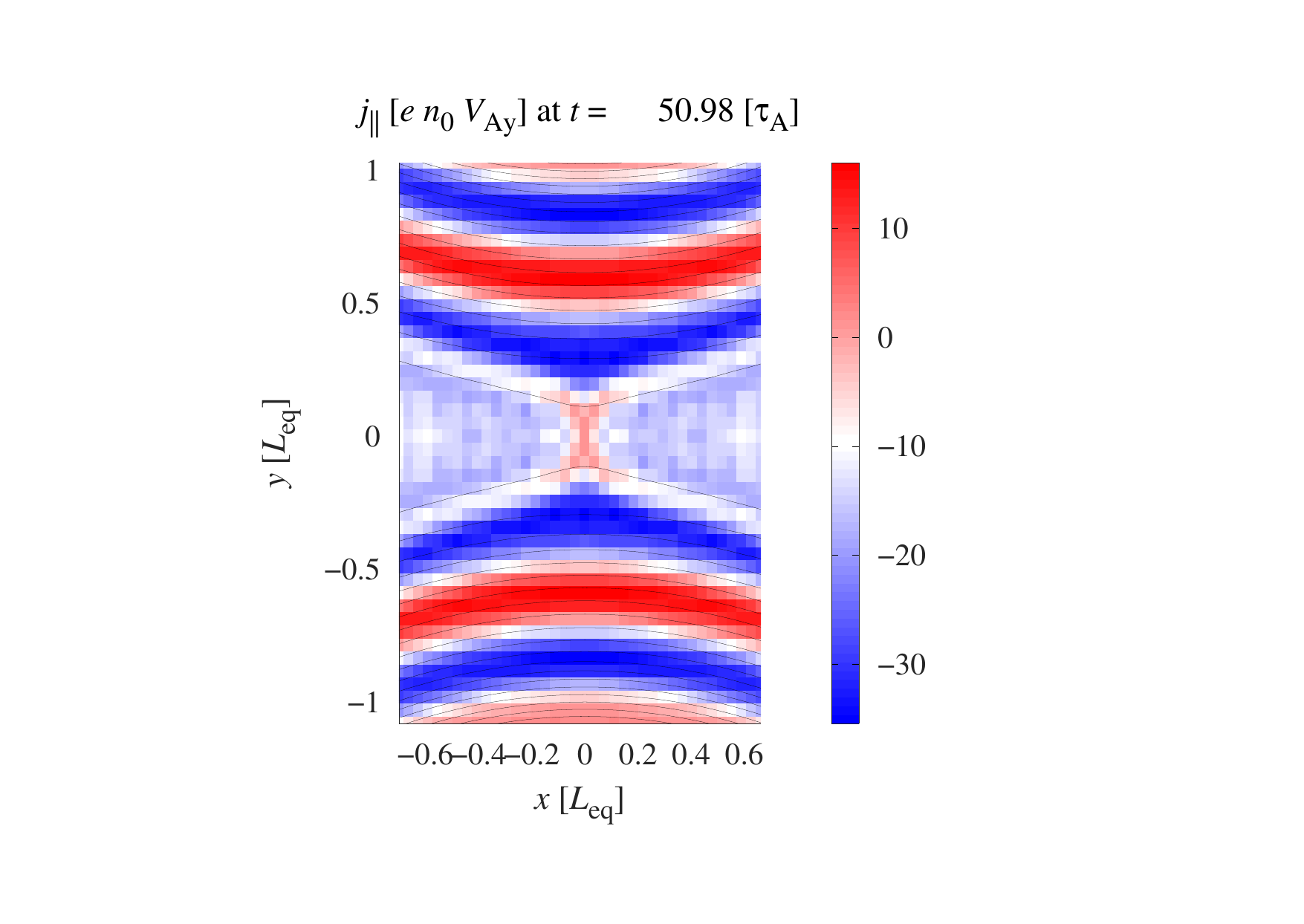}
    \caption{Top: Contour of the parallel current density $u_e=U_e$ for simulation $4_{F}$  ($1728^2$). Bottom:  Contour of the parallel current density $j_{\parallel}$ of simulation $4_{GK2}$.  Isolines of the magnetic potential are superimposed on all the color maps. }
    \label{fig:case4contour}
\end{figure}

Figure \ref{fig:case4contour} shows the evolution of the secondary current sheet for the cases $4_{GK3}$ (lowest $\beta_e$ gyrokinetic case of this series) and $4_F$. The current sheet formed in the two frameworks does not follow the separatrices but remains mainly aligned along $x=0$. 

This comparison makes it possible to show that, by simply adding bi-fluid effects resulting from a large sonic Larmor radius, one can switch from a marginally stable case to a marginally unstable case. This result, shown by \cite{Gra2022pl} by means of the fluid model, is thus confirmed by gyrokinetic simulations.

\section{Energy partition - Similarities and differences between gyrokinetics and gyrofluid} \label{ssec:compar}
\subsection{Energy components}

As we consider here a plasma with no collisions, the gyrokinetic system solved by {\tt AstroGK} conserves the total energy (Hamiltonian) \citep{How06,Sch09}, normalized by $B_{0}^2/(4\pi)$

\begin{equation}
\label{gyroenergy} 
    W(\delta {\mathcal F}_{e},\delta {\mathcal F}_{i})
    = \frac{1}{2}
    \int \td x \td y \left(
    \frac{\tau_s\beta_e}{2} \sum_s \frac{1}{n_0} \int \td \hwu \hcalf \jo \delta {\mathcal F}_{s}^2  + |\nabla_{\perp} \apar|^2  +  d_i^2 |\bpar|^2 \right)
\end{equation}
where 
$\delta {\mathcal F}_{s}=g_{s} + (1/di) (2Z_{s}/(\tau_s \beta_e)) (\jo^2-1) \phi + d_i \vp^2 \ju\jo \bpar$ is the perturbation of the particle distribution function for the species $s$. The first term is the perturbed entropy of the species $s$, while the second term and third terms are the energy of the perpendicular and parallel perturbed magnetic field. 
We can extract the first two moments from the perturbed particle distribution function as

\begin{equation} \label{decomp}
\delta {\mathcal F}_{s} = d_i n_s + \sqrt{\frac{2\sigma_s}{\tau_s\beta_e}} d_i \vpar u_s + h_s',
\end{equation}
where the perturbed density and parallel velocity of the particle of species $s$ are denoted as $n_s$ and $u_s$, respectively, and $h_s'$ contains all higher moments of the perturbed distribution function. By definition,
\begin{align}
    \int \td \hwu \hcalf \jo h_s' & = 0, &
    \int \td \hwu \hcalf \vpar \jo h_s' & = 0.
\end{align}

We can therefore decompose the expression (\ref{gyroenergy}) in the following way

\begin{equation} \label{gyroenergy_decomp}
    W(\delta {\mathcal F}_{e}, \delta {\mathcal F}_{i}) =
    \frac{1}{2}
    \int \td x \td y \left(
    \sum_s
    \left( \tau_s \rho_s^2 n_s^2
    + \sigma_s d_i^2 u_s^2
    + \frac{\tau_s\beta_e}{2} \frac{1}{n_0} \int \td \hwu \hcalf \jo {h_s'}^2
    \right)
    + |\nabla_{\perp} \apar|^2
    + d_i^2 |\bpar^2|
    \right).
\end{equation}

The first term is the energy generated by the electron density variance, the second term is the kinetic energy of the parallel electron flow, and the third term is the free electron energy.\\
With regard to the collisionless gyrofluid model, the system of equations (\ref{conteiso}) - (\ref{ampperpcondiso}) possesses a conserved Hamiltonian given by
 \begin{align}
 &H_{gf}(N_e , A_e )=\frac{1}{2} \int dx dy \, \left( \rho_s^2 N_e^2  + d_e^2 U_e^2 + | \nabla_\perp^2 \apar |^2   - N_e ( \gamue \phi - \rho_s^2 2 \gamde \bpar)\right).  \label{ham2f}
 \end{align}
We remark that, as will be shown in the Appendix, the form of the Hamiltonian (\ref{ham2f}), obtained from the quasi-static closure, is the same that one obtains by imposing what we refer to as an isothermal gyrofluid closure (the relations between $\phi, \apar, \bpar $ and $N_e, U_e$ will, however, be different in the two cases).
 
 Using the relation (\ref{ne2}) and (\ref{ue2}) we can also write the Hamiltonian in terms of particle variables as follows:
 \begin{align}
 H_{p}(n_e ,A_e )= & \frac{1}{2} \int dxdy \, \left( \rho_s^2 n_e \gamue^{-2} n_e  + d_e^2 \left( \gamue^{-1}u_e \right)^2 + | \nabla_\perp^2 \apar |^2  +
  d_i^2 |\bpar|^2 \right. \nno \\
  & \left. + n_e \left( 1 - 2\gamue^{-2} \right)\phi + \phi \left( \gamue^{-2}-1 \right) \frac{\phi}{\rho_s^2} \right). 
  \label{ham2fp}
 \end{align}

When we consider the limit $\bee$, $m_e/m_i \rightarrow 0$ the Hamiltonian of the gyrofluid equations is reduced to 
 \begin{align}
 \label{Hfluid}
 &H_p(n_e , A_e)=\frac{1}{2} \int dxdy \, \left( \rho_s^2 n_e^2 + d_e^2 u_e^2 + | \nabla_\perp^2 \apar|^2  + |\nabla_{\perp} \phi|^2  \right), 
 \end{align}
 which is namely the Hamiltonian of the fluid Eqs. (\ref{fluid1}) - (\ref{fluid2}). 
In Eq. (\ref{Hfluid}), the contribution from left to right are the energy generated by the electron density fluctuation, the parallel electron kinetic energy, the perpendicular magnetic energy and the perpendicular plasma kinetic energy which is essentially the $\textbf{E} \times \textbf{B}$ flow energy.

\subsection{Negligible $\beta_e$: fluid vs gyrokinetic} 

\begin{figure}
    \centering
\includegraphics[trim=0 0 0 0, scale=0.35]{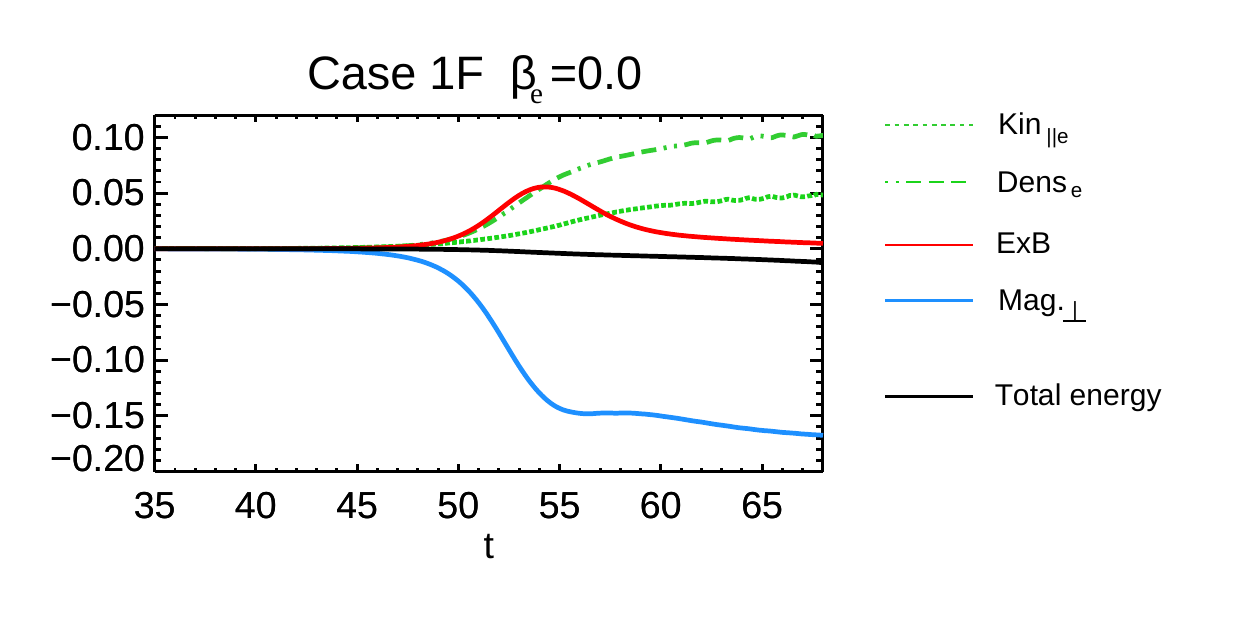}\includegraphics[trim=0 0 0 0, scale=0.7]{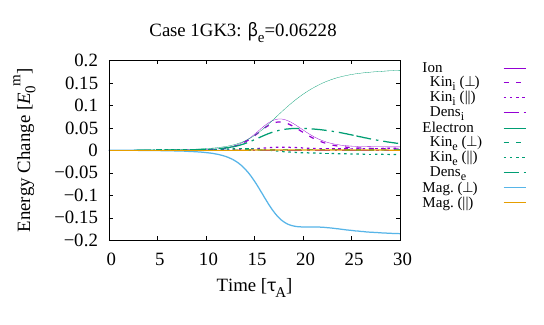}\\
\includegraphics[trim=0 0 0 0, scale=0.42]{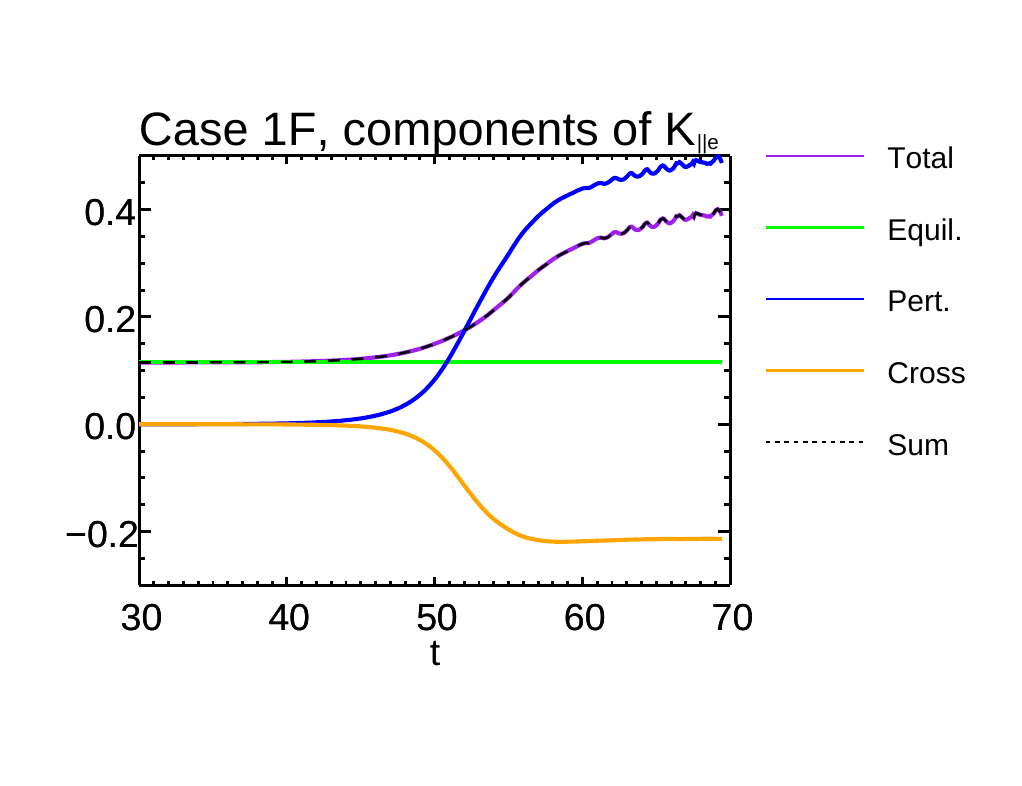}\includegraphics[trim=0 0 0 0, scale=0.57]{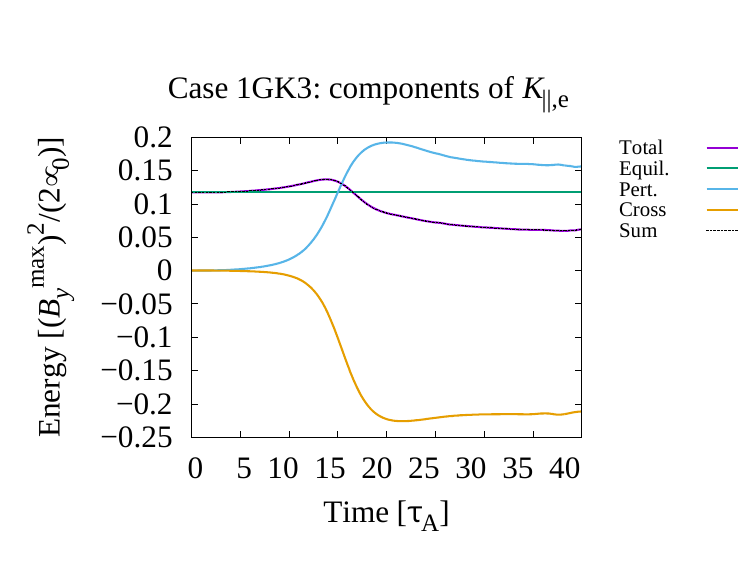}
      \caption{Top: Time evolution of the energy variations for the cases $1_F$ (left) resolution and $1_{GK3}$ (right). Bottom: Change of the parallel kinetic energy for the same simulations. No plasmoid in this case.}
    \label{fig:energycase1}
\end{figure}

On Fig. \ref{fig:energycase1} we present the comparison between the energy variation of the fluid case $1_F$ and that of the low $\beta_e$ gyrokinetic case $1_{GK3}$ ($\beta_e=0.062$). The variations are defined as $(1/2)\int \td \br( \xi(x,y,t) - \xi(x,y,0)) / E(0)$ where the function $\xi$ can be replaced by the different contributions to $\hat{W}$ and $H$ (where $\hat{W}$ is also considered in the 2D limit)  and $E(0)$ is the initial total energy.
On the gyrokinetic plots, the four main energy channels are shown as solid lines. The solid purple line is the total ion energy variation. We also show the evolution of the variations relative to the density variance (dashed dotted), the parallel kinetic energy (densely dashed) and the perpendicular kinetic energy (loosely dashed), that are components of the total particle energy. The same channels are shown for the electrons in green. 

The amount of magnetic energy that is converted is identical between fluid and gyrokinetics and appears to be transferred mainly to the electrons. On the other hand, it is not identically distributed in the gyrokinetic and fluid frameworks. For the fluid simulations, the magnetic energy has no choice but to be converted into electron density fluctuations or electron parallel acceleration, whereas in the gyrokinetic case, there is little energy sent to these channels. This suggests that, in the gyrokinetic framework, the energy of the electrons increases due the fluctuations of the higher order moments of the distribution function due to phase mixing (\cite{Lou13, Num15}), such as for instance, the perpendicular and parallel electron temperature. It is likely that the magnetic energy is actually converted into thermal electron energy. Such possibility is prevented in the fluid case because, as a consequence of the closure, for $\beta_e \rightarrow 0$, no temperature fluctuations are allowed.

The striking difference between the two approaches is that the parallel electron cinetic energy increases in the fluid case, whereas it is quasi-constant or decreasing in the gyrokinetic one (Fig. \ref{fig:energycase1}). In order to investigate the origin of this difference, we  performed an initial condition check and decomposed the parallel electron kinetic energy. The decomposition leads to three energy components, namely the equilibrium part ($u_{eq}^2$), the perturbation part ($\tilde{u}_e^2$) and the cross term ($2\tilde{u}_eu_{eq}$). The change of each component is shown on the bottom panel of Fig. \ref{fig:energycase1}. The equilibrium contribution clearly does not change in time. The quadratic perturbation part is always positive but globally the variation of parallel electron kinetic energy can decrease because of the cross term becoming negative, which is the case for the gyrokinetic simulation. For the fluid case, the perturbation term increases considerably, leading to a positive variation of the parallel kinetic energy, since the electrons are highly accelerated for conservation of the total energy.\\
%

 With regard to the ions, the closure assumptions imply an even rougher approximation of the ion dynamics, in the fluid case, with respect to gyrokinetics. In the gyrokinetic case, for low $\beta_e$, we can see on Fig. \ref{fig:energycase1} that the main component of the total ion energy consists of the perpendicular kinetic energy, which is included in the ${h_s}'$ part in Eq.~\eqref{gyroenergy_decomp}\footnote{Since the $\textbf{v}_{\perp}$ moments are generally not orthogonal, we cannot clearly separate each of them.}, given by 
\beq  \label{ionperp}
\frac{1}{2} \int \td x \td y d_i^2 \textbf{u}_{\perp,i} ^2.
\eeq

where the perpendicular ion velocity $\textbf{u}_{\perp , i}$ is calculated directly from its definition as a moment in the following way:

\begin{align}
d_i \textbf{u}_{\perp,i} = \sqrt{\frac{\tau_i\beta_e}{2\sigma_i}}\frac{1}{n_{0}} \int \td \hwu {\mathcal F}_{eq_i} \textbf{v}_{\perp} \delta {\mathcal F}_{i}
\end{align}

Notice that the perpendicular flow holds the identity from the definition
\begin{equation}
    d_i \textbf{u}_{\perp,i} = (-\nabla \phi - \rho_s^2 \nabla \cdot \texttt{p}_{\perp\perp,i})
    \times \textbf{z}
\end{equation}
where the perpendicular pressure tensor is given by
\begin{align}
d_i \texttt{p}_{\perp\perp,i} =  \tau_i \frac{1}{n_0} \int \td \hwu {\mathcal F}_{eq_i}
\textbf{v}_{\perp} \textbf{v}_{\perp} \delta {\mathcal F}_{i}
\end{align}
The perpendicular flow is given by the sum of $\textbf{E}\times\textbf{B}$ drift and diamagnetic drift of  perturbed pressure.

In spite of the closure, the evolution of the energy component (\ref{ionperp}) is very similar to that of the $\textbf{E}\times\textbf{B}$ flow energy of the gyrofluid case. 
For a very small $\beta_e$, no parallel ion kinetic energy and parallel magnetic energy seems to be generated.

\subsection{Finite $\beta_e$: gyrofluid vs gyrokinetic}

\begin{figure} 
    \centering
\includegraphics[trim=0 0 0 0, scale=0.32]{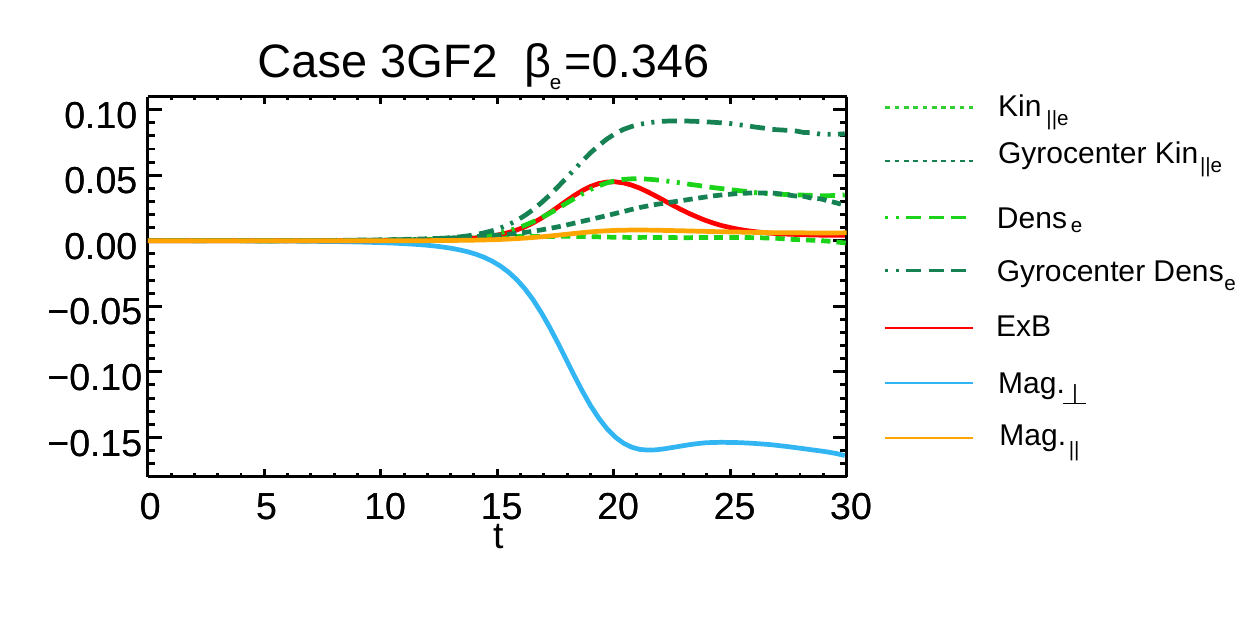}\includegraphics[trim=0 0 0 0, scale=0.7]{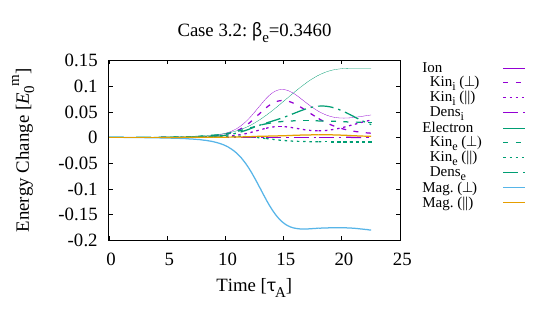}\\
\includegraphics[trim=0 0 0 0, scale=0.32]{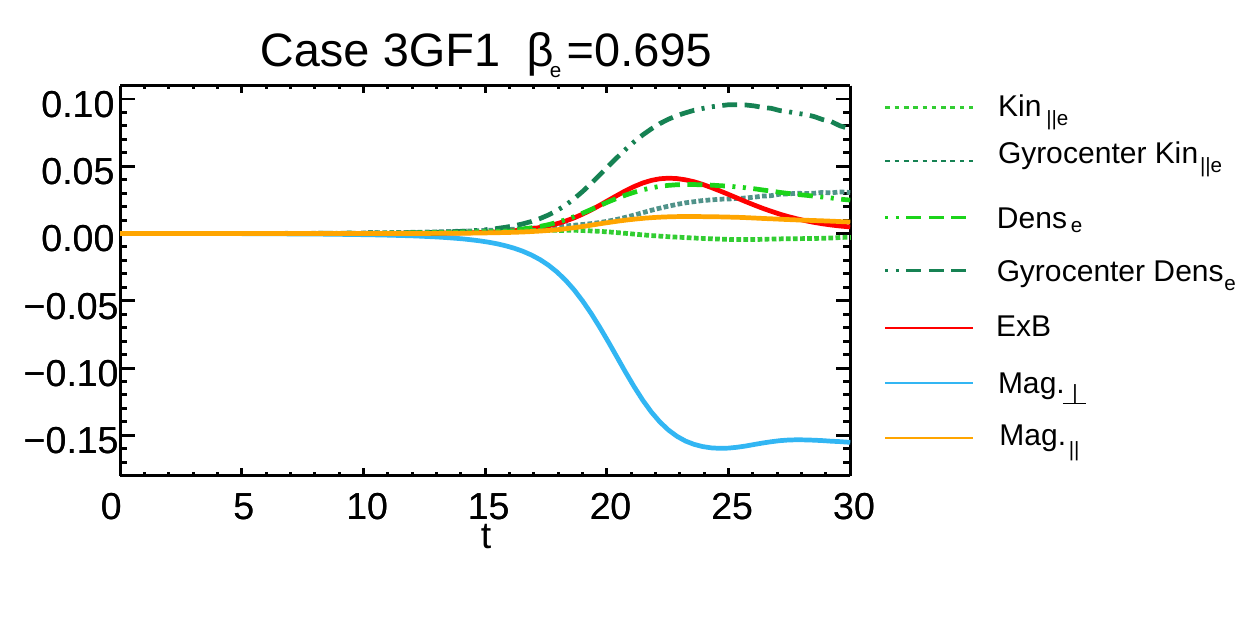}\includegraphics[trim=0 0 0 0, scale=0.7]{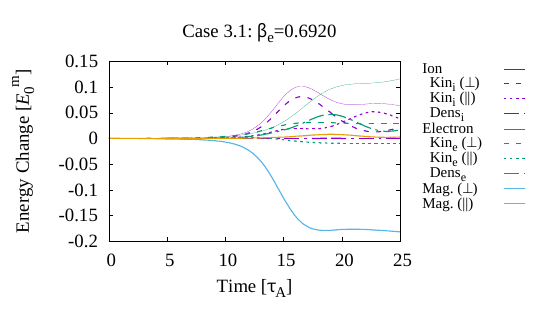}
      \caption{Time evolution of the energy variations for the cases $3_GF$ and $3_{GK}$. }
    \label{fig:energycase3}
\end{figure}

 When $\beta_e$ is very small, the FLR corrections become negligible and the particle and gyrocenter variables coincide. On the other hand, for non-negligible $\beta_e$, the electron Larmor radius becomes finite and the relations (\ref{ne}) and (\ref{ue}) allow us to relate the density and parallel velocity of the particles to those of the gyrocenters. On Fig. \ref{fig:energycase3} we compare the gyrofluid energy variations with the gyrokinetic ones for $0 < \beta_e < 1$. For this purpose, we use the simulation set for $p=3$.

In the plot referring to the gyrofluid energy, we show the variation of both the particles and gyrocenters energy. For instance, the curve referring to "Kin$_{\parallel e}$" corresponds to the variation of $(1/2)\int dxdy d_e^2 u_e^2$, which is comparable to the second term of the gyrokinetic energy (\ref{gyroenergy_decomp}).  The one referring to "Gyrocenter Kin$_{\parallel e}$" corresponds to the variation of $(1/2)\int dxdy d_e^2 U_e^2$. By increasing $\beta_e$, the difference between the variation of the energy of the gyrocenters and that of the particles broadens. 
With finite $\beta_e$, we now note a loss of parallel kinetic energy of the electrons for the gyrofluid case, which is in better agreement with the gyrokinetic approach. Increasing $\beta_e$, will also generate more parallel magnetic energy, which is well reproduced by the gyrofluid model. 
On the other hand,  the gyrokinetic cases indicate that a significant part of the magnetic energy is now converted into parallel ion kinetic energy. As already mentioned, a limitation of the reduced gyrofluid model is that the ion parallel velocity has been "artificially" removed by imposing $U_i=u_i=0$. The limitations of this assumption become evident, in particular, from Fig. \ref{fig:energycase3} which shows that, in the gyrokinetic case, for sufficiently large $\beta_e$, the ion fluid is actually accelerated along the $\bz$ axis. On the other hand, it seems that despite this missing element, the gyrofluid model is suitable for studying the formation of plasmoid for $0<\beta_e<1$.

\section{Conclusion}

In this work, we have numerically investigated the plasmoid formation employing both gyrofluid and gyrokinetic simulations, assuming a finite, but small $\beta_e$. This made it possible to show that the effect of finite $\beta_e$, associated to finite electron Larmor radius effects, promotes the plasmoid growth. These results contribute to shed light on collisionless reconnection mediated by the plasmoid instability, and in particular on the role of the effects present at the electron scale.  

 This work showed the ability of the reduced gyrofluid model to achieve relevant new insights into current-sheet stability and magnetic reconnection. In particular, predictions on marginal stability on current sheets, obtained by \cite{Gra2022pl} in the fluid limit, were confirmed by gyrokinetic simulations. It also indicates that the fluid and gyrofluid models make it possible to obtain accurate results in short computational times.

The comparison between the gyrofluid and the gyrokinetic models reveals key similarities and differences between the two frameworks, which gives insight into the important underlying physical effects. Indeed, the adopted gyrokinetic model is a $\delta f$ model from which the gyrofluid model can be derived with appropriate approximations and closure hypotheses. This allowed to directly identify possible limitations of the closures applied to the gyrofluid moments, that distinguish the gyrofluid model from its gyrokinetic parent model. We therefore presented the impact of the closure on the distribution and conversion of energy during reconnection. The closure, which does not allow for parallel temperature fluctuations, implies that the energy must be converted into fluctuations of density and parallel velocity of the electrons. This is not in agreement with the gyrokinetic simulations, but does not seem to interfere with the formation of plasmoids. In particular, for relatively small but finite $\beta_e$, the hypothesis of absent parallel ion motion made in the gyrofluid framework is valid and does not affect the plasmoid instability. The gyrokinetic perpendicular ion velocity is well represented by the fluid $\textbf{E}\times \textbf{B}$ velocity. On the other hand, gyrokinetic simulations show a large fraction of magnetic energy transferred to fluctuations of higher order moments.

\subsection*{Acknowledgements}

The work of RN was partly supported by JSPS KAKENHI Grant Number 22K03568.
The numerical simulations were performed using the EUROfusion high performance computer Marconi Fusion and Galileo100 hosted at CINECA (project FUA35-FKMR and IsC86 MR-EFLRA), the computing facilities provided by the Mesocentre SIGAMME hosted by the Observatoire de la C\^ote d'Azur, as well as JFRS-1 supercomputer system at the Computational Simulation Centre of the International Fusion Energy Research Centre (IFERC-CSC) at the  Rokkasho Fusion Institute of QST (Aomori, Japan) and on the 'Plasma Simulator' (NEC SX-Aurora TSUBASA) of NIFS with the support and under the auspices of the NIFS Collaboration Research program (NIFS22KISS019).

\subsection*{Appendix. Comparison between the gyrofluid isothermal and the quasi-static closure}   \label{app:energy}

In this Appendix we first show how the gyrofluid Hamiltonian (\ref{ham2f}) can be obtained from the gyrokinetic Hamiltonian (\ref{gyroenergy}) by applying the quasi-static closures and the assumptions described in Sec. \ref{ssec:connect}. Subsequently, we compare with the gyrofluid Hamiltonian obtained by applying what we refer to as gyrofluid isothermal closure.

The conserved energy (\ref{gyroenergy}) of the $\delta f$ gyrokinetic model used for the comparison in this paper can be expressed in terms of the gyrocenter perturbed
distribution function

\begin{equation}
    \hfs = \delta {\mathcal{F}}_s + \frac{1}{d_{i}} \frac{2 Z_s}{\tau_s \beta_e} \left(\phi - \left\langle \phi -
    d_i \sqrt{\frac{\tau_s\beta_e}{2\sigma_s}}\bv_{\perp}\cdot\textbf{A}_{\perp} \right\rangle_{\textbf{X}_s}\right),
\end{equation}

where $\langle \ \rangle_{\textbf{X}_s}$ denotes the gyroaverage at constant guiding center $\textbf{X}_s$ and $\textbf{A}_\perp$ is the magnetic vector potential associated with parallel magnetic perturbations, so that $\nabla \times \textbf{A}_{\perp} = \bpar \mathbf{z}$. In this way we obtain

\begin{align}
    W(\delta {\mathcal F}_{e},\delta {\mathcal F}_{i}) = H_{gy}(g_{e},g_{i}) & =
    \frac{1}{2} \int \td x \td y
    \sum_{s} \frac{1}{n_{0}} \int \td \hwu \hcalf
    \left(
    \frac{\tau_{s}\beta_{e}}{2} g_{s}^{2} \right. \nno \\
    & 
    \left.
    + \frac{Z_s}{d_{i}} g_{s}
    \left( \jo \phi + \sqrt{\frac{\tau_s\beta_e}{2\sigma_{s}}} \vpar \jo \apar
    + \frac{\tau_s}{Z_s} \rho_s^2 \vp^2 \ju \bpar\right)
    \right)
    \label{hamdfs}
\end{align}

Note that, in Eq. (\ref{hamdfs}), we already took the spatial 2D limit, in order to directly obtain the gyrofuid Hamiltonian.

The gyrocenter perturbed distribution function can be developed as a series of its gyrocenter moments using Hermite and Laguerre polynomials. Here, we will retain only the first two moments of the hierarchy for the two species, and apply two different closures, namely the quasi-static closure \cite{Tas20}, and a gyrofluid isothermal closure, where the perpendicular and parallel gyrocenter temperature fluctuations $T_{\perp s } = T_{\parallel s}=0$, are set equal to zero (as all the other higher order gyrofluid moments). Such kind of closure is applied, for instance, by \cite{S_2010}, although in this Reference, gyrocenter temperature, as well as heat flux fluctuations, are retained and all the other higher order moments are set equal to zero.

For the quasi static closure, the expansion of the gyrocenter perturbed distribution functions for the two species are given by 

\begin{align}
g_{e}
 & = d_i N_{e} + \sqrt{\frac{2\sigma_e}{\beta_e}} d_i \vpar U_{e} 
 - \sum_{n=1}^{\infty}
 L_{n}\left(\frac{\vp^2}{2}\right)
 \left(G_{1ne} \frac{1}{d_i} \frac{2Z_e}{\beta_e} \phi + 2 G_{2ne} d_{i} \bpar\right)
\label{dfeqs},
\\
g_{i}
& = - \sum_{n=1}^{\infty} L_{n} \left( \frac{\vp^2}{2} \right)
\left( G_{1ni} \frac{1}{d_{i}} \frac{2}{\tau_i\beta_e} \phi
+ 2 G_{2ni} d_{i} \bpar \right)
\label{dfiqs}
\end{align}

The difference between electron and ion treatments in Eqs. (\ref{dfeqs}) and (\ref{dfiqs}), is clearly due to the assumption (\ref{ioncond}). We mention that, by retaining, in Eq. (\ref{dfiqs}), also ion gyrocenter density and parallel velocity fluctuations, and applying the same procedure described in the following, one can derive the energy of the 4-field Hamiltonian gyrofluid model described by \cite{Gra22flr}.

In the case of the gyrofluid isothermal closure the truncated expansion simply gives\\

\begin{align}
g_{e}
 & = d_i N_{e} + \sqrt{\frac{2\sigma_e}{\beta_e}} d_i \vpar U_{e},
\\
g_{i}
& = 0.
\end{align}

To simplify the infinite sums in (\ref{dfeqs}) and (\ref{dfiqs}), we can make use of the following relations \citep{Sze75}

\begin{align}
J_{0} (\alpha_s) & =
e^{-b_s/2} \sum_{n=0}^{\infty} \frac{L_n\left(\vp^2/2\right)}{n\text{!}}
\left( \frac{b_s}{2} \right)^n, \\
2 \frac{J_{1}(\alpha_s)}{\alpha_s} & = 
e^{-b_s/2} \sum_{n=0}^{\infty} \frac{L_n^{(1)}\left(\vp^2/2\right)}{(n+1)\text{!}}
\left( \frac{b_s}{2} \right)^n,
\end{align}

where $L_n^{(1)}$ are associated Laguerre polynomials and of the expression of the operators (\ref{G1s})-(\ref{G2s}).
We therefore obtain the following equality, that can be injected in (\ref{dfeqs}) and (\ref{dfiqs}),

\begin{align}
    \sum_{n=1}^{\infty}
    L_{n} \left(\frac{\vp^{2}}{2}\right)
    \left( G_{1ns} \frac{1}{d_i} \frac{2Z_{s}}{\tau_s\beta_e} \phi + 2 G_{2ns} d_i \bpar \right)
    & =
    \left( J_{0}(\alpha_s) - G_{10s}\right) \frac{1}{d_i} \frac{2Z_s}{\tau_s\beta_e} \phi
    +
    \left( \vp^2 \frac{J_{1}(\alpha_s)}{\alpha_s} - 2G_{20s} \right) d_i \bpar.
    \label{rela}
\end{align}

We can now reduce the gyrokinetic Hamiltonian (\ref{hamdfs}) to gyrofluid ones by injecting the two sets of truncated perturbed gyrocenter distribution functions.  In the quasi-static case, thanks to the relation (\ref{rela}), all the contributions involving $G_{1n_s}$ and $G_{2n_s}$, with $n \geq 1$, cancel. It turns out that the two  resulting gyrofluid Hamiltonians can be written in an identical form, which corresponds to the following
\begin{align}
 H( N_e , A_e ) & =
 \frac{1}{2} \int \td x \td y \, \left(   \rho_s^2  N_e^2 - A_e \lue ( A_e)  - N_e ( \gamue \call_\phi ( N_e)  \right.  \nno\\
 & ~~~ \left.  - \rho_s^2 2 \gamde \call_B ( N_e))\right), \label{ham2fder}
 \end{align}

where we recall that  $A_e=\gamue \apar - d_e^2 U_e$ and the linear operators $\lue$, $\call_\phi$ and $\call_B$ are given by
 \begin{align}
 &\bpar=\call_B ( N_e), \qquad \phi=\call_\phi ( N_e),\\
 & U_e=\lue ( A_e).
 \end{align}
through the quasi-neutrality relation and the two components of Amp\`ere's law.
The expression (\ref{ham2fder}) coincides, up to integration by parts, to the Hamiltonian (\ref{ham2f}).
 
Evidently, the quasi-static  quasineutrality  equation and Amp\`ere's law will differ from the isothermal ones. Therefore, the total conserved energy are actually evolving differently and the operators $\call_B$, $\call_{\phi}$, $\call_{U_e}$ are closure-dependent operators.

For instance, the explicit form of the quasi-neutrality relation, in the quasi-static case, writes
\begin{align}
  - \gamue N_e + \lapp \phi + (\gamue^2 -1) \frac{\phi}{\rho_s^2} + (1 -\gamue 2 \gamde)\bpar  =0, \label{qnqs}
\end{align} 
while in the case of the gyrofluid isothermal closure, we have
\beq
- G_{10e} N_e + \lapp \phi+ ( \Gamma_{0e} - 1 ) \frac{\phi}{\rho_s^2}+ (1 - \Gamma_{0e} - \Gamma_{1e} ) \bpar =0, \label{qniso}
\eeq
and where the $\Gamma_{0,1 e}$ operators are defined in Fourier space in the following way
\beq \label{gammaoperatore}
\Gamma_{0e}  \rightarrow  I_0\left(\kpe \frac{\beta_e}{2}d_e^2\right) e^{-\kpe \frac{\beta_e}{2}d_e^2} , \qquad \Gamma_{1e}   \rightarrow I_1\left(\kpe \frac{\beta_e}{2}d_e^2\right) e^{-\kpe \frac{\beta_e}{2}d_e^2},
\eeq
where $I_n$ are the modified Bessel functions of order $n$.\\

A first comment is that the two closures, and consequently the two sets of static equations are in fact identical if we assume
\beq 
G_{1ne} = G_{2ne} = 0, \quad \text{for}  \quad n \geq 1,
\eeq
 which gives the approximations
\beq  \label{approxoper}
\Gamma_{0e}(b_e)^{1/2} = G_{10e}, \qquad (\Gamma_{0e} (b_e)-\Gamma_{1e}(b_e))^{1/2} = 2 G_{10e} G_{20e}.
\eeq
On the other hand, the relations (\ref{approxoper}) can be interpreted in a different way, i.e. considering the exact expressions (\ref{gammaoperatore}) for $\Gamma_{0e}$  and $\Gamma_{1e}$, and assuming that the expressions for $G_{10e}$ and $G_{20e}$ can be adapted to match the quasi-neutrality relations following from the two closures. In this way, from the first relation in Eq. (\ref{approxoper}), one retrieves, for the case $s=e$, the approximate expression for the operators $G_{10s}$ introduced by \cite{Dor93}. The advantages of this approach have been more recently discussed also by \cite{Man18}.
The above approach is indeed reminiscent of the approach used by \cite{Dor93} in order to find an expression for $G_{10s}$ yielding a better agreement of liner gyrofluid theory with the linear gyrokinetic theory. A similar approach, accounting also for $G_{20s}$ in a finite-$\beta$ gyrofluid model, was followed by \cite{Des11}.

A second point is that, in the limit of negligible ion and electron Larmor radius, when considering  $\tau_i \rightarrow 0$ and  $\beta_e \rightarrow 0$, the two sets of static relations become identical and we obtain the same fluid Hamiltonian 

For instance, both (\ref{qniso}) and (\ref{qnqs}) will reduce to the quasi-neutrality relation
\beq
N_e  = \lapp \phi,
\eeq
which is going to give rise to the $\textbf{E}\times\textbf{B}$ flow  energy in the fluid Hamiltonian.

\nocite{*}

\bibliographystyle{plainnat}

\bibliography{Granier_plasmoids}

\end{document}